\documentclass[%
superscriptaddress,
 preprint,
showkeys,showpacs,preprintnumbers,
amsmath,amssymb,
aps,
prb,
]{revtex4-2}

\usepackage{graphicx}
\usepackage{dcolumn}
\usepackage{bm}

\usepackage{upgreek}
\usepackage{amsmath}

\begin{document}


\title{Flat optics for analog computing: from fundamental mechanisms to advanced meta-processors}

\author{Tingting Liu}
\email{ttliu@ncu.edu.cn}
\affiliation{School of Information Engineering, Nanchang University, Nanchang 330031, China}
\affiliation{Institute for Advanced Study, Nanchang University, Nanchang 330031, China}

\author{Jumin Qiu}
\affiliation{School of Physics and Materials Science, Nanchang University, Nanchang 330031, China}

\author{Xintong Shi}
\affiliation{School of Information Engineering, Nanchang University, Nanchang 330031, China}

\author{Qiegen Liu}
\affiliation{School of Information Engineering, Nanchang University, Nanchang 330031, China}

\author{Shuyuan Xiao}
\email{syxiao@ncu.edu.cn}
\affiliation{School of Information Engineering, Nanchang University, Nanchang 330031, China}
\affiliation{Institute for Advanced Study, Nanchang University, Nanchang 330031, China}

\begin{abstract}
As the explosive growth of visual data increasingly strains the latency and energy limits of conventional  electronic computing, optical analog computing has re-emerged as a disruptive paradigm for zero-power, speed-of-light information processing. Propelled by the unprecedented wave-manipulation capabilities of optical metasurfaces, this field is undergoing a rapid transition from macroscopic physical optics to ultra-compact, on-chip meta-processors. This Review examines the fundamental mechanisms of metasurface-empowered optical computing spanning Fourier-domain, nonlocal spatial-domain, and interferometric architectures that perform mathematical operations, with a particular focus on spatial differentiation and edge detection as representative computing tasks. By emphasizing recent breakthroughs, we highlight the evolution of meta-processors from static, linear regimes to dynamically reconfigurable, nonlinear, and quantum-assisted multidimensional platforms. We also envision how the synergy of AI-driven inverse design and the integration of analog meta-front-ends with optical neural networks will synergistically revolutionize next-generation intelligent machine vision.

\end{abstract}

\keywords{optical analog computing, metasurfaces, spatial differentiation, edge detection, nonlocal flat optics, reconfigurable meta-optics, optical neural networks}
\maketitle


\section{\label{sec1}Introduction}

Driven by the explosive growth of visual data in the era of artificial intelligence, big data, and autonomous driving, conventional electronic computing hardware is increasingly bottlenecked by processing latency and high energy consumption. To satisfy the relentless demand for higher computing power, optical computing has emerged as a disruptive paradigm. Utilizing photons as information carriers, optical systems offer unprecedented bandwidth, low crosstalk, and inherent massive parallelism. While optoelectronic hybrid computing architectures have been extensively developed to bridge the ultra-high speed of optics with the programming flexibility of electronics\cite{Shastri2021,McMahon2023,Shekhar2024}, fully unleashing the potential of light for specific continuous-data tasks requires a more direct approach. Consequently, all-optical analog computing, which processes information directly in its native analog form at the speed of light without analog-to-digital conversion, has resurged as a highly promising solution. Operating with minimal energy dissipation, optical analog computing is exceptionally suited for high-throughput front-end machine vision tasks, such as edge detection. However, conventional optical analog signal processors, such as the classic 4$f$ spatial filtering systems, rely heavily on bulky refractive lenses, fundamentally hindering their integration into modern compact nanophotonic circuits.

To circumvent the dimensional restrictions of conventional optics, optical metasurfaces have emerged as a revolutionary platform for all-optical computing. Metasurfaces are two-dimensional planar arrays of subwavelength artificial scatterers, named as meta-atoms that offer unprecedented degrees of freedom to manipulate the amplitude, phase, polarization, and dispersion of electromagnetic waves \cite{Yu2014,Chen2016,Hu2021a,Fu2024}. Fundamentally, both electrical and optical analog computing can be characterized by the mathematical framework of transfer function engineering. Just as a digital filter relies on a kernel function, the operational characteristics of an optical computing system are determined by its optical transfer function (OTF). Metasurfaces uniquely enable the precise customization and engineering of these transfer functions at the nanoscale. By judiciously tailoring the geometry, size, and spatial arrangement of meta-atoms, these ultra-thin planar devices can encode complex mathematical transfer functions directly into their local or non-local optical responses. Pioneered by the concept of "computational metamaterials" introduced by Silva et al. in 2014 \cite{Silva2014}, metasurfaces successfully map specific mathematical operators, such as spatial differentiation into physical structures. This multidimensional modulation capability not only eliminates the bulky footprint of conventional optical systems but also allows specific mathematical operations to be executed effortlessly as the light beam traverses them.

\begin{figure*}[htbp]
	\centering
	\includegraphics[width=\linewidth]{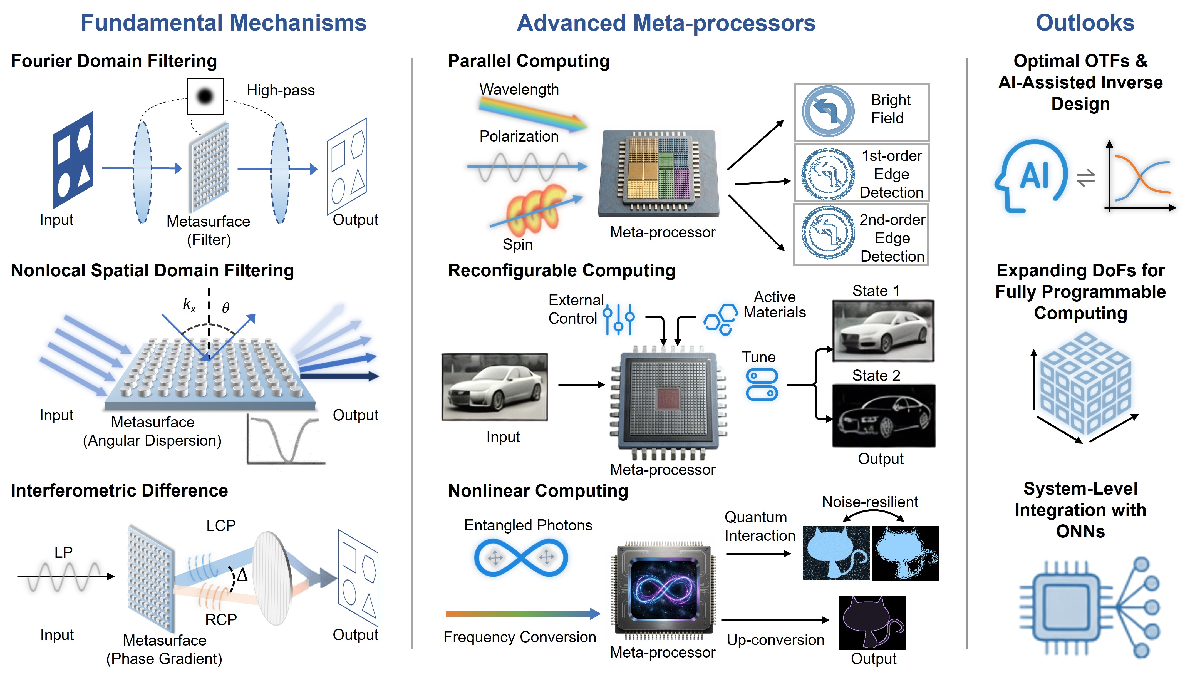}
	\caption{Schematic overview of our review in the development of metasurface-empowered optical analog computing. The field progresses from fundamental physical mechanisms, spanning Fourier-domain filtering, nonlocal spatial-domain filtering, and interferometric difference, toward advanced meta-processors capable of parallel multiplexing, dynamic reconfigurability, and nonlinear computing. The future outlook envisions AI-assisted inverse design for optimal optical transfer functions (OTFs), degrees of freedom (DoFs) expansion for fully programmable optical computing,  and the system-level integration of optical analog front-ends with optical neural networks (ONNs). }
	\label{fig1}
\end{figure*}

This Review aims to provide a comprehensive and up-to-date overview of the rapid progress in metasurface-empowered optical analog computing, placing a distinct emphasis on spatial differentiation and image edge detection. While previous literature has extensively covered early computational metasurfaces \cite{Abdollahramezani2018,ZangenehNejad2020Analogue,Wu2022,Badloe2022,He2022,Xu2022, Wang2024a,Zhou2024a,Zhou2025All,Cui2026}, this review uniquely spotlights the recent multidimensional paradigm shifts towards  functional integration, practical, and advanced meta-processors, such as computing under ambient incoherent illumination, quantum-entangled edge detection, and direct angular spectrum manipulation. As shown in Fig. \ref{fig1} , it is structured as follows: Section 2 elucidates the fundamentals of transfer function engineering, establishing the mathematical frameworks and physical mapping mechanisms, along with their extensions to phase objects and angular spectrums. Section 3 systematically categorizes the physical implementations in flat optics into three primary architectures: Fourier-domain systems, nonlocal spatial-domain devices, and interferometric difference platforms. Subsequently, Section 4 explores the cutting-edge trajectory of advanced meta-processors, highlighting the transition from static, single-task operations toward parallel multiplexed, dynamically reconfigurable, and nonlinear regimes. Finally, Section 5 concludes by addressing current physical trade-offs and outlines future perspectives on AI-assisted inverse design and the system-level integration of optical analog front-ends with optical neural networks.

\section{\label{sec2} Fundamental principles of transfer function engineering}

To design an all-optical analog computing system, the essential prerequisite is to establish a rigorous mapping between the desired mathematical operators and the physical modulation of electromagnetic waves. Inspired by linear systems theory, this mapping can be elegantly described through the lens of transfer function engineering. In this section, we elucidate the fundamental mathematical framework of linear shift-invariant (LSI) systems, detail the mathematical models governing spatial differentiation, and systematically classify the physical implementation strategies based on their transfer function modulation domains.

\subsection{\label{sec2.1} Mathematical framework}

In the context of optical analog computing, an optical system or a metasurface can be mathematically treated as a black box acting on an input optical field $f(x,y)$. The output optical field $g(x,y)$ is determined by the two-dimensional (2D) spatial convolution of the input signal and the system's spatial impulse response $h(x,y)$. By applying the Fourier transform, this real-space convolution simplifies to a direct multiplication in the spatial frequency domain, i.e., momentum space or $k$-space, as:
\begin{equation}\label{eq1}
G(k_{x},k_{y})=H(k_{x},k_{y}) F(k_{x},k_{y}),
\end{equation}
where $k_{x}$ and $k_{y}$ represent the transverse spatial frequency variables.  $F(k_{x},k_{y})$ and $G(k_{x},k_{y})$ are the spatial Fourier transforms of the input and output fields, respectively. Crucially, $H(k_{x},k_{y})=\mathcal{F}\{{h(x,y)}\}$ is defined as the optical transfer function (OTF) of the system. Therefore, the core objective of spatial analog computing is to physically design a flat optical structure whose transmission or reflection coefficient precisely matches the ideal OTF $H(k_{x},k_{y})$ associated with the target mathematical operator. It is worth noting that for vectorial electromagnetic waves, the input and output optical fields are treated as vectors comprising transverse electric (TE) and transverse magnetic (TM) components. Consequently, the OTF is generalized into a $2\times2$ Jones matrix tensor, affording additional degrees of freedom to perform mathematical operations via polarization coupling.

In image processing and machine vision, edge detection relies on identifying abrupt variations in the intensity or phase of an image, which is mathematically equivalent to computing the spatial derivative of the optical field. Based on the derivative property of the Fourier transform, spatial differentiation corresponds to multiplying the signal by its spatial frequency in the Fourier domain. Specifically, the $n^{th}$ derivative of a two-dimensional wavefunction, into which an arbitrary function can be decomposed, is related to the corresponding wavenumber and the one-dimensional spatial Fourier transform of that function as follows:
\begin{equation}\label{eq2}
\frac{\partial^n f(x, y)}{\partial x^n} = \mathcal{F}^{-1}\left\{(ik_x)^n \mathcal{F}\{f(x, y)\}\right\},
\end{equation}
where $i=\sqrt{-1}$ . Given that the field expansion is based on a continuum of wavenumbers, the transfer function corresponding to $n^{th}$-order differentiation follows a parabolic profile as the order $n$ increases. 

To perform a first-order spatial differentiation along a single axis, e.g., the $x$ axis, the operation is defined as $g(x)=\partial f(x)/\partial x$. In the Fourier domain, the ideal OTF takes the form of $H(k_{x})\propto ik_{x}$. Mathematically, this is an odd function, dictating two stringent physical requirements: the amplitude response must drop to zero at normal incidence where $k_{x}=0$, and there must be a sharp $\pi$ phase jump crossing the origin. Implementing a first-order differentiator fundamentally necessitates breaking the spatial mirror symmetry of the optical structure or utilizing oblique incidence.

For the second-order spatial differentiation $g(x) = \partial^{2} f(x) / \partial x^{2}$ , the corresponding OTF in the $k$-space is proportional to $(ik_{x})^2$, yielding: $H(k_x) \propto -k_{x}^{2}$.  To achieve isotropic 2D edge detection, the Laplacian operator is widely employed as $\mathcal{F}\{\nabla^2 f(x, y)\} = -(k_{x}^{2} + k_{y}^{2}) \mathcal{F}\{f(x, y)\}$. The 2D OTF is consequently expressed as $H(k_{x}, k_{y}) \propto -(k_{x}^{2} + k_{y}^{2})=-k_{||}^{2}$, where $k_{||}$is the in-plane wavevector magnitude. In contrast to the first-order derivative, the second-order OTF is an even function featuring a parabolic amplitude profile with respect to the spatial frequency, reaching its minimum (ideally zero) at normal incidence where $k_{x}=k_{y}=0$. Because it possesses an even symmetry and requires a relatively constant phase profile across the bandwidth, the second-order differentiation is physically more straightforward to implement. It can be naturally synthesized by exploiting symmetric optical resonances, such as Fano resonances or bound states in the continuum at normal incidence without the need to break the structural symmetry.

\subsection{\label{sec2.2} Physical mapping mechanisms}

Translating the aforementioned mathematical models into physical nanophotonic devices requires a rigorous mapping between the spatial frequencies ($k_{x}, k_{y}$) of the desired transfer function and the tangible optical parameters of the computational system. Based on where and how the transfer function is manipulated, the prevailing implementation strategies can be categorized into three mainstream approaches: Fourier domain filtering, spatial domain filtering, and optical interferometric difference.

\subsubsection{\label{sec2.2.1} Fourier domain filtering}

This approach is inspired by conventional 4$f$ correlator systems \cite{Silva2014}. When an incident optical field with field of $f(x, y)$ passes through a Fourier lens, the transverse wavevectors ($k_{x},k_{y}$) are physically mapped to spatial coordinates ($x,y$) on a focal plane. Here, a metasurface is engineered with a spatially varying transmittance $T(x,y)$ to physically emulate $H(k_{x},k_{y})$. Then a Fourier lens can be used for the realization of an inverse Fourier transform, at the expense of image mirroring of the desired output with the relation of  $\mathcal{F} ^{-1}\{\mathcal{F}\{f(x, y)\}\} \propto f(-x, -y)$. Mathematically, when an incident optical field $f(x,y)$ passes through the first lens with focal length $f_{l}$, the transverse wavevectors ($k_{x},k_{y}$) of the input signal are projected onto the spatial coordinates ($x_{f},y_{f}$) at the back focal plane according to the relation of $x_f = \frac{f_{l}}{k_{0}} k_{x}$$,$ $ y_f = \frac{f_{l}}{k_{0}} k_y$, where $k_{0}=2\pi/ \lambda_{0}$ is the free-space wavevector. By designing a planar metasurface with a position-dependent complex transmission or reflection coefficient $T(x_{f}, y_{f})$, the target optical transfer function $H(k_{x},k_{y})$ is physically encoded into the spatial profile of the metascreen, $T(x_f,y_f) \equiv H \left( \frac{k_0 x_f}{f_l}, \frac{k_0 y_f}{f_l} \right)$. Subsequently, the second metalens performs an inverse Fourier transform, yielding the final output field: $g(x,y) = \mathcal{F}^{-1} \{ T(x_f,y_f) \mathcal{F} \{f(x,y)\} \}$. 
 
The prominent advantage of the Fourier filtering approach is its capability to implement arbitrarily complex and high-order transfer functions, as the spatial profile $T(x_f, y_f)$ can be highly customized pixel-by-pixel. However, the necessity of cascaded components including input plane, processing mask, and output plane, hinders its ultimate miniaturization and monolithic integration into ultra-compact nanophotonic circuits. Recently, single-lens computational systems and highly compact meta-imagers have been developed to condense the Fourier transform and OTF modulation into a single-layer structure \cite{Fu2022,Wang2022}, significantly advancing the integration of Fourier domain filtering.

\subsubsection{\label{sec2.2.2} Spatial domain filtering}

To obviate the need for coordinate transformation lenses, spatial domain filtering that is historically referred to as the Green's function approach, implements the desired transfer function directly in real space \cite{Silva2014,Cotrufo2023a}. This is achieved by exploiting the spatial nonlocality, i.e., angular dispersion of a uniform metasurface or resonant thin film.

Any arbitrary incident field profile $f(x,y)$ can be decomposed into a continuum of linearly polarized plane waves including TE and  TM via spatial Fourier transform. The transverse wavevector of each plane wave component is intrinsically linked to its angle of incidence $\theta$ relative to the surface normal, as $k_{||}=\sqrt{k_{x}^{2} + k_{y}^{2}}=k_{0} \sin{\theta}$. Instead of using lenses to separate these components, the approach engineers the  the spatial nonlocality, usually angular dispersion of the metasurface. The structure's transmission or reflection coefficient naturally exhibits an angle-dependent response, which acts directly as the mathematical kernel $H(k_{x},k_{y})$. Considering the vectorial nature of electromagnetic waves, this physical mapping is generally described by a $2\times2$ transfer function tensor $\bar{H}(k_{x},k_{y})$ that couples different polarization states:
\begin{equation} \label{eq3}
    \begin{pmatrix} G_{TE}(k_x, k_y) \\ G_{TM}(k_x, k_y) \end{pmatrix} = \begin{pmatrix} H_{TE-TE}(k_x, k_y) & H_{TE-TM}(k_x, k_y) \\ H_{TM-TE}(k_x, k_y) & H_{TM-TM}(k_x, k_y) \end{pmatrix} \begin{pmatrix} F_{TE}(k_x, k_y) \\ F_{TM}(k_x, k_y) \end{pmatrix},
\end{equation}
where the on-diagonal elements represent the coupling of parallel polarizations, and off-diagonal elements represent orthogonal polarization conversions. The output processed signal in real space is directly obtained as the waves exit the structure, 
\begin{equation}\label{eq4}
g(x, y) = \iint \bar{H}(k_x, k_y) F(k_x, k_y) e^{i(k_x x + k_y y)} dk_x dk_y.
\end{equation}

To tailor $\bar{H}(k_{x},k_{y})$ to fit desired mathematical kernels, various optical phenomena are leveraged. These are generally categorized into resonance-based mechanisms  such as Fano resonances, surface plasmon polaritons (SPPs), and guided-mode resonances, and non-resonance-based mechanisms such as the Brewster effect or the spin Hall effect of light. Mathematically, the connection between the desired mathematical kernels and the physical resonances of metasurfaces is established through the Taylor series expansion. For instance, in a non-resonance-based or resonance-based system, such as the Brewster effect or surface plasmon polaritons, the reflection or transmission coefficient $R(k_{x})$ can be designed to drop to zero at a specific incidence angle with $k_{x}=0$ in the local coordinate. Expanding $R(k_{x})$ around this zero point yields,  $R(k_x) \approx R(0) + R'(0) k_x + \frac{1}{2} R''(0) k_x^2 + O(k_x^3)$. Since $R(0)=0$, the dominant term becomes linearly proportional to $k_{x}$ (i.e., $R(k_{x}) \propto ik_{x})$, inherently acting as a first-order differentiator for signals with a narrow spatial bandwidth. Similarly, systems with even symmetry evaluated at normal incidence naturally eliminate the first-order term, leaving the parabolic $k_{x}^{2}$ term, which seamlessly executes the second-order  differentiation operation. While the spatial domain filtering, also referred to Green's function approach offers a true ultra-thin, single-element computing platform, realizing an arbitrary complex transfer function remains challenging due to the inherent constraints of physical dispersion relations.

\subsubsection{\label{sec2.2.3} Optical interferometric difference}

An alternative to direct frequency filtering is the interferometric difference approach, which captures the variation between discrete quantities. Mathematically, the derivative $f'(x)$ can be approximated using the central finite difference method:
\begin{equation}\label{eq5}
f'(x) \approx \frac{f(x+h)-f(x-h)}{2h},
\end{equation} 
where $h$ signifies a tiny spatial displacement. A prominent physical mechanism for this is polarization multiplexing utilizing the photonic spin Hall effect (SHE) and Pancharatnam-Berry (PB) phase \cite{Zhou2019b,Zong2023}. By arranging anisotropic meta-atoms with a spatially varying orientation angle $\theta(x,y)$, the metasurface imparts opposite geometric phases ($\pm 2\theta$) to the left-handed (LCP) and right-handed circularly polarized (RCP) components. When illuminated by a linearly polarized beam, this spin-dependent phase gradient induces a slight transverse spatial shift $\Delta$ between the LCP and RCP components. If an analyzer is placed orthogonally to the incident polarization after the metasurface, the transmitted output field $E_{\text{out}}$ becomes the coherent superposition or subtraction of the two shifted beams. Mathematically, this finite difference tightly approximates the spatial derivative:
\begin{equation}\label{eq6}
E_{\text{out}}(x, y) \propto E_{\text{in}}(x + \Delta, y) - E_{\text{in}}(x - \Delta, y) \approx 2\Delta \frac{\partial E_{\text{in}}(x, y)}{\partial x}.
\end{equation}
This approach completely bypasses the need for stringent phase-matching conditions or narrow resonant bandwidths, offering a highly efficient and broadband solution for isotropic 1D and 2D edge detection.

\subsection{\label{sec2.3} Extension of transfer functions}

The concept of transfer function engineering can be expanded beyond simple amplitude objects and 2D spatial dimensions to accommodate more complex computing tasks.

\subsubsection{\label{sec2.3.1} Spiral phase filtering}

While the aforementioned derivative models effectively extract boundaries based on intensity variations, they are primarily tailored for amplitude objects. However, in many critical scenarios, such as biomedical imaging, living cells and unlabelled biological tissues are highly transparent and barely absorb light. These targets are classified as phase objects, as they only alter the phase delay of the transmitting wave rather than its amplitude. For a thin phase object, the transmitted optical field can be approximated as $E_{in}(x,y) = A_0 \exp[i\Phi(x,y)] \approx A_0[1 + i\Phi(x,y)]$, where $\Phi(x,y)$ represents the spatial phase retardation caused by local refractive index or thickness variations. Traditional edge detection relies on converting these invisible phase gradients into observable intensity contrasts, which conventionally requires bulky interferometric setups. To address this in a compact analog computing framework, spiral phase contrast has emerged as a powerful solution. By introducing an optical vortex with a helical phase profile into the momentum space, one can achieve isotropic, two-dimensional edge enhancement for both amplitude and phase objects \cite{RitschMarte2017,Huo2020,Kim2022}.

In the spatial frequency domain, the OTF of a spiral phase filter is expressed as, $H(k_x,k_y)=\exp(i\varphi_k)$, where $\varphi_k = \arctan(k_y/k_x)$ is the azimuthal angle in the $k$-space. To understand its mathematical core, we can compare it to a 1D spatial differentiator. As discussed previously, a 1D first-order derivative requires a $\pi$ phase jump across the origin, which is analogous to a 1D Hilbert transform characterized by sign$(k_{x})$. Remarkably, the spiral phase function  $\exp(i\varphi_k)$ intrinsically provides exactly a $\pi$ phase difference across the center along any arbitrary diametric line. Therefore, applying $\exp(i\varphi_k)$ as a Fourier filter serves as an isotropic 2D generalization of the Hilbert transform, this is also mathematically known as the Riesz transform. It effectively applies a directional derivative isotropically to all radial directions around the center.

\subsubsection{\label{sec2.3.2} Spatiotemporal transfer functions}

While spatial analog computing processes the transverse spatial frequencies $(k_x,k_y)$, modern optical processing increasingly demands the manipulation of ultrafast temporal signals. In the time domain, a temporal differentiator operates on the complex envelope $e_{in}(t)$  of an optical pulse with a carrier frequency $\omega_0$. Based on the Fourier transform of a time-domain signal, the ideal optical transfer function $H(\omega)$ for an $n^{th}$ order temporal differentiation in the frequency domain is expressed as: $H(\omega) = \frac{E_{out}(\omega)}{E_{in}(\omega)} \propto [E(\omega - \omega_0)]$. Physically, this requires a metamaterial or a resonant grating that exhibits zero transmission or reflection exactly at the carrier frequency $\omega_0$ , accompanied by an abrupt $\pi$ phase jump. Recently, by simultaneously tailoring the spatial nonlocality and temporal dispersion, corresponding to the angular dispersion and frequency dependence of a single metasurface, researchers have pioneered spatiotemporal transfer functions $H(\omega,k_x,k_y)$ \cite{Rajabalipanah2021,Zhou2023a,Zhou2023}. Such devices can execute mixed derivatives $\frac{\partial^2}{\partial x \partial t}$, opening new frontiers for event-based vision and ultrafast moving-target detection.

\subsubsection{\label{sec2.3.3} Angular spectrum transfer functions}

While traditional optical analog computing manipulates spatial frequencies to differentiate the spatial profile of an image, recent innovations have pushed the boundaries to perform mathematical operations directly on the angular spectrum itself \cite{Deng2024,Deng2024a}. In this novel paradigm, rather than filtering the spatial domain, the ideal optical transfer function is engineered to perform differentiation with respect to the transverse wavevectors, $\partial / \partial k_x$ or $\partial^2 /\partial k_x \partial k_y$. This unique mathematical extension isolates and enhances extremely weak angular spectrum features from strong overlapping backgrounds, establishing a powerful theoretical framework for derivative spectroscopy and advanced trace analysis. The broadband physical implementation of this concept will be further discussed in Section 4.1.

\section{\label{sec3} Physical implementations in flat optics}

The paradigm of all-optical analog computing using subwavelength artificial structures was fundamentally established by the seminal work of Silva et al. in 2014 \cite{Silva2014}. By introducing the concept of "computational metamaterials," this pioneering research demonstrated that ultra-thin metamaterial blocks could be tailored to perform mathematical operations, such as spatial differentiation, integration, and convolution, directly on the profile of an impinging electromagnetic wave. To physically synthesize the desired OTFs, the authors proposed two distinct architectural approaches, laying the very foundation for what we now categorize as Fourier domain filtering and spatial domain filtering.

\subsection{\label{sec3.1} Fourier domain filtering systems}

The physical realization of arbitrary complex OTFs is deeply rooted in early theoretical explorations of nanoscale optical antennas and screens. Preceding and concurrent with the formalization of "computational metamaterials", Monticone et al. proposed the concept of composite metascreens, demonstrating full control over nanoscale optical transmission \cite{Monticone2013}. In the same year, Farmahini-Farahani et al. laid crucial theoretical groundwork by demonstrating that 2D arrays of subwavelength optical nanoantennas could be meticulously engineered to process optical signals \cite{FarmahiniFarahani2013}. Building upon these discrete scatterer concepts, Silva et al. conceptualized the first highly miniaturized version of the classic $4f$ spatial filtering system \cite{Silva2014}. Termed the "Metasurface (MS) approach", this architecture performs mathematical operations in the spatial Fourier domain through indirect wavevector modulation by sandwiching a subwavelength structured metascreen between two graded-index (GRIN) dielectric waveguides which act as compact Fourier transformers. By precisely engineering the inhomogeneous distribution of the metamaterial blocks, the metascreen physically encodes the ideal OTF. While revolutionary in replacing traditional bulky lenses, this approach fundamentally relied on the physical separation of spatial and frequency domains. 

\begin{figure*}[htbp]
	\centering
	\includegraphics[width=\linewidth]{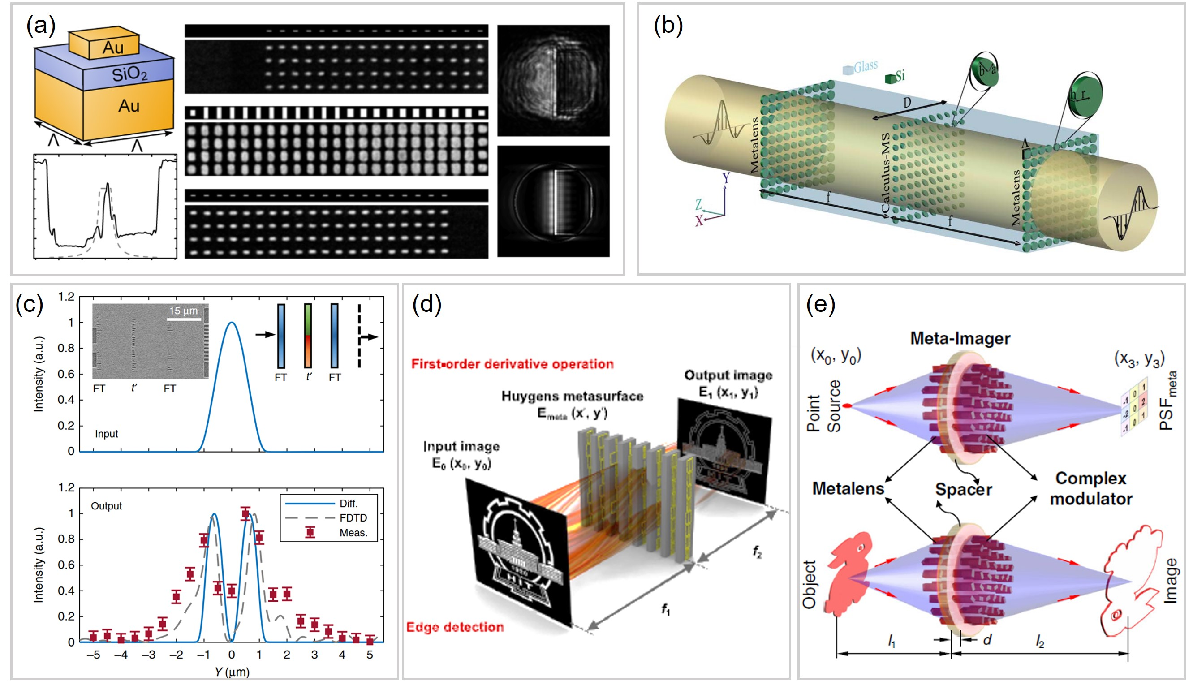}
	\caption{ Representative architectures for Fourier-domain filtering in optical analog computing. (a) A reflective metasurface utilizing gap-surface plasmon resonances for spatial differentiation \cite{Pors2015}. (b) A transmissive all-dielectric metasurface based on electric and magnetic resonances \cite{Abdollahramezani2017}. (c) An ultra-compact, single-layer high-contrast transmitarray metasurface on a silicon-on-insulator substrate for on-chip differentiation \cite{Wang2019}. (d) A single-layer Huygens’ metasurface designed for independent phase and amplitude modulation, facilitating basic mathematical operations such as spatial differentiation and cross-correlation without cascading lenses \cite{Wang2022}. (e) An ultracompact meta-imager integrating an image-forming metalens with a complex-amplitude meta-modulator, enabling arbitrary and functionality-unlimited kernels for all-optical convolutional computing \cite{Fu2022}.}
	\label{fig2}
\end{figure*}

Following the foundational transmissive $4f$ architecture, subsequent research rapidly evolved to optimize the physical footprint and operational flexibility. A significant structural optimization was the transition from transmissive to reflective configurations. Pors et al. pioneered this by demonstrating analog computing using gap-surface plasmon metasurfaces in reflection mode shown in Fig. 2(a) \cite{Pors2015}. This geometry elegantly halves the physical footprint of the computing system, as the same lens or GRIN slab serves simultaneously as both the Fourier and inverse Fourier transform modules. Shortly after, Abdollahramezani et al. expanded this filtering paradigm into a compact configuration by integrating graphene-based metalines, where the total length of the whole structure including lenses and metalines is about $\lambda/4$, about 60 times shorter than the one proposed by Silva et al. \cite{AbdollahRamezani2015}. This also introduced a crucial degree of freedom of dynamic tunability, allowing the OTF to be actively reconfigured by local tuning of the chemical potential of each unit cell to achieve any transmission phase and amplitude profile. 

While early plasmonic metasurfaces offered excellent subwavelength confinement, they suffered fundamentally from inherent Ohmic absorption losses. To overcome this critical bottleneck, researchers transitioned to all-dielectric platforms. Chizari et al. successfully demonstrated a reflective analog computing architecture based on a high-efficiency dielectric meta-reflect-array(MRA) which consists of engineered nanobricks on top of a silica spacer and a silver substrate \cite{Chizari2016}. It supports both electric and magnetic dipolar Mie-type modes, as well as multiple reflections within the spacer layer, enabling fully manipulating the amplitude, and the phase profile of the reflected cross-polarized light spatially by varying nanoresonator dimensions. This approach significantly mitigated energy dissipation while maintaining high-fidelity spatial differentiation. Beyond basic calculus, the mathematical capabilities of 4$f$-type metasurfaces were further elevated to solve complex mathematical models. Abdollahramezani et al. demonstrated that by meticulously engineering the spatially varying phase and amplitude profiles of dielectric metasurfaces in the Fourier plane as Fig. 2(b) shows, the system could be configured to solve first-order integro-differential equations and second-order ordinary differential equations, showcasing the immense potential of metasurfaces as versatile optical analog solvers \cite{Abdollahramezani2017}.

Concurrently, researchers explored unconventional meta-atom topologies to push analog computing into more challenging spectral regimes. Chen et al. successfully extended optical spatial differentiation into the visible wavelength spectrum \cite{Chen2017}. By employing a unique silver dendritic metasurface, they leveraged tailored resonances to precisely impart the required $\pi$ phase jump for first-order derivatives. This meta-atom design strategy was later expanded by this group to synthesize the singular transfer functions required for 2D spatial integration in the visible regime \cite{Chen2020}.

While the aforementioned $4f$ systems offer unparalleled flexibility in tailoring complex OTFs, their inherent requirement for spatial propagation to physically separate spatial frequencies poses a persistent barrier to miniaturization.  For compactness, Wang et al. presented a one-dimensional high-contrast transmitarray metasurface shown in Fig. 2(c), and control the coherent inference of the parallelly transmitted wavefront with low loss by slightly adjusting the dimension of the void slots defined in the device layer \cite{Wang2019}. By performing spatial differentiation and integration on guided modes within an on-chip dielectric metasurface, they completely bypassed the volumetric constraints of free-space $4f$ systems, paving the way for ultra-compact, chip-scale optical processors. To further compress the computational footprint, researchers sought to execute operations without relying on far-field diffraction. Bao et al. addressed this by designing transmissive metasurfaces capable of performing mathematical operations directly in the near-field through simultaneous and independent modulation of subwavelength phase and amplitude \cite{Bao2020}. This drive to eliminate free-space propagation ultimately converges with the pursuit of fully integrated nanophotonics, without the need of blocks for making Fourier transform or Fourier lens.   

Recent breakthroughs have successfully condensed these multi-stage architectures into highly compact, single-lens or single-layer meta-devices, fundamentally merging coordinate transformation and OTF modulation. A prime example of this architectural compression was demonstrated by Wang et al., who proposed a single-layer spatial analog meta-processor utilizing a Huygens' metasurface in Fig. 2(d) \cite{Wang2022}. By algorithmically superimposing a focusing phase factor with the desired mathematical kernel, this single-layer structure compresses a bulky $4f$ system into a highly compact $2f$ configuration. The carefully designed Huygens' meta-atoms achieve independent, full-coverage modulation of both transmission amplitude and phase without the need for auxiliary Fourier lenses, enabling precise 2D edge detection and complex cross-correlation operations. Operating on a similar principle of volumetric compression, Fu et al. developed an ultracompact "meta-imager" designed for arbitrary all-optical convolutional computing \cite{Fu2022} in Fig. 2(e). This system ingeniously integrates an image-forming metalens with a complex-amplitude meta-modulator into a single device. Instead of applying a filter mask in a physically separated Fourier plane, the meta-modulator is designed to directly shape the point spread function (PSF) as the spatial domain equivalent of the OTF. This architecture enables massively parallel convolutional operations in real-time, effectively bridging the gap between rigorous spatial Fourier filtering and the extreme dimensional constraints of modern machine vision systems. Yang et al. disigned a compact, flexible, and precise optical analog computing kernel for pattern recognition by scheme of optical analog cross-correlation computing implemented via a dielectric metasurface, where the Fresnel diffraction process is approximated as a Fourier transform \cite{Yang2025a}. This design further enhances system integration. Together, these single-lens and single-layer meta-architectures represent a paradigm shift in Fourier-based optical computing. By merging the phase profiles required for coordinate transformation  such as imaging and focusing, with the complex amplitude modulation required for mathematical filtering in OTF engineering, they effectively bridge the gap between the rigorous mathematical capabilities of spatial Fourier filtering and the extreme dimensional constraints of modern machine vision systems.

\subsection{\label{sec3.2}Nonlocal spatial domain filtering systems}

To achieve true chip-scale integration, the computational architecture must be shifted entirely into the real space. Spatial domain filtering achieves this by exploiting the inherent angular dispersion of a flat optical device. Rather than physically separating the angular components of the beam, this approach executes mathematical operations by engineering the nonlocal angular dispersion of a flat device. Because an incident beam's spatial frequencies $(k_x,k_y)$ are strictly mapped to its angles of incidence, the angle-dependent transmission or reflection profile of a surface can act directly as the mathematical kernel $H(k_x,k_y)$. By directly mapping the OTF onto the angular dispersion of the system, this architecture eliminates the need for Fourier lenses entirely, achieving ultra-compact, real-space wave-based computing and setting the theoretical baseline for subsequent nonlocal flat optics.

\subsubsection{\label{sec3.2.1}Non-resonant mechanisms}

\begin{figure*}[htbp]
	\centering
	\includegraphics[width=\linewidth]{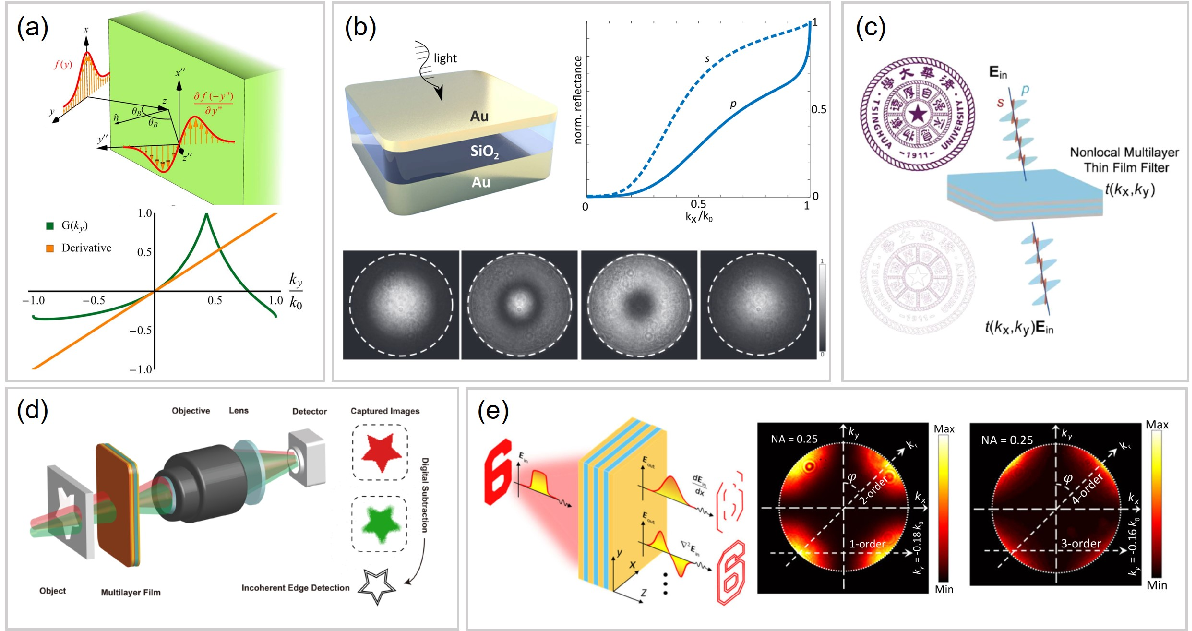}
	\caption{ Non-resonant platforms for nonlocal spatial-domain filtering in analog computing. (a) Schematic illustration of an unstructured dielectric interface performing first-order spatial differentiation based on the Brewster effect\cite{Youssefi2016}. (b) A highly efficient spatial high-pass filter utilizing a Salisbury screen configured as a thin-film perfect absorber \cite{Wesemann2019}. (c) A transmissive nonlocal multilayer thin-film filter engineered for isotropic 2D image differentiation \cite{Jin2021}. (d) Optimized multilayer films demonstrating highly robust spatial differentiation under ambient incoherent optoelectronic illumination\cite{Zhang2022}. (e) A tailored planar dielectric multilayer chip capable of executing multiple-order (third- and fourth-order) analog spatial differentiation \cite{Liu2022}.}
	\label{fig3}
\end{figure*}

The earliest and most elegant realizations of spatial domain filtering relied on the intrinsic angular dispersion found in fundamental optical phenomena and basic thin-film structures. Silva et al. concurrently proposed the "Green's function approach" that performs mathematical operations by engineering the nonlocal angular dispersion of a multilayered metamaterial slab. By operating away from narrow geometric resonances, these mechanisms generally offer higher robustness, broader bandwidths, and drastically simplified fabrication. Following this, researchers realized that even simpler, naturally occurring macroscopic optical phenomena inherently possess the exact angular dispersion required for basic calculus. 

A quintessential example is the Brewster effect, which Youssefi et al. brilliantly repurposed for analog computing \cite{Youssefi2016}. At the Brewster angle, the reflection coefficient for a TM polarized plane wave drops precisely to zero, as shown in Fig. 3(a). By performing a Taylor series expansion around this zero-reflection point, a simple, unstructured dielectric interface naturally acts as an ultra-broadband first-order spatial differentiator.  Building upon the elegance of 1D single-interface phenomena, researchers rapidly sought to expand the dimensionality of these optical solvers. Zangeneh-Nejad and Khavasi demonstrated that a simple dielectric slab could perform spatial integration \cite{ZangenehNejad2017}. Shortly after, to overcome the intrinsic polarization dependence of the Brewster effect, which is strictly limited to TM-polarized waves, Zangeneh-Nejad et al. utilized a half-wavelength dielectric slab at a specific oblique incidence angle to realize highly efficient, polarization-independent first-order spatial differentiation \cite{ZangenehNejad2018}. Furthermore, by replacing the dielectric slab with a plasmonic graphene film, they not only achieved extreme miniaturization (approximately $\lambda/100$) but also endowed the differentiator with dynamic reconfigurability by electrically tuning its chemical potential.  While the 1D Brewster effect offers striking simplicity, its fundamental reliance on the $p$-polarized component along the plane of incidence limits it to 1D differentiation. To break this dimensional barrier, Xu et al. successfully advanced the zero-reflection concept to two dimensions \cite{Xu2020}. Rather than relying on a structured metasurface, they ingeniously harnessed the spin-orbit interaction of light at an isotropic air-glass interface. By meticulously analyzing the cross-polarization coupling combined with the Brewster anomaly, they achieved fully isotropic 2D spatial differentiation, elegantly proving that even basic optical interfaces possess untapped multidimensional computing potential.

While single interfaces or homogenous slabs are brilliantly simple, their angular bandwidth (Numerical Aperture, NA) is often restricted, limiting the resolution of the processed images. To achieve more precise control over the OTF, researchers turned to multilayer thin films. A pioneering demonstration in this regime was presented by Wesemann et al., who engineered a near-perfect absorbing thin-film mirror based on a Salisbury screen configuration in Fig. 3(b) \cite{Wesemann2019}. By tailoring this ultrathin Au/SiO$_2$/Au multilayer structure to completely absorb normally incident light ($k_x=k_y=0$) while strongly reflecting higher spatial frequencies, the mirror natively acts as a highly efficient spatial high-pass filter, selectively lighting up the edges of both amplitude and phase images.  The design of multilayer differentiators was significantly advanced by Wu et al. \cite{Wu2017}. By combining the transfer matrix method with optimization algorithms, they inversely designed Si/SiO$_2$ multilayer stacks to achieve the parabolic OTF required for second-order differentiation. Notably, this inverse-design framework is highly flexible, allowing the architectures to accommodate varying spatial bandwidths. 

Building upon this multilayer concept, subsequent research offered vast degrees of freedom via transfer-matrix optimization. Jin et al. leveraged this approach to design transmissive nonlocal multilayer optical filters in Fig. 3(c), eliminating the bulky beam-splitters required in reflection-based setups \cite{Jin2021}. Such thin film filter can be fabricated using mature wafer-scale thin film deposition technique, and exhibit an optimized nonlocal optical response, for isotropic image differentiation in transmission mode for arbitrary input polarization. To push the resolution limits, Xue et al. utilized advanced optimization algorithms to design multilayer films capable of high-NA optical edge detection, significantly broadening the operational angular bandwidth to capture finer image details \cite{Xue2021}. Furthermore, a profound leap toward practical, real-world biological imaging was achieved by Zhang et al. \cite{Zhang2022}. Historically, optical analog computing strictly required coherent laser illumination to maintain phase relationships, which suffers from severe speckle noise and limits compatibility with standard microscopes. Zhang's team successfully optimized multilayer films in Fig. 3(d) to perform spatial differentiation under incoherent optoelectronic illumination. By relaxing the strict requirement for spatial coherence, this incoherent differentiator drastically improves image quality and paves the way for seamless integration into conventional bright-field microscopy.

As the physical mechanisms for tailoring angular dispersion mature, the frontier of spatial analog computing has shifted from executing single tasks to performing multiple independent operations simultaneously. Abdolali et al. and Babaee et al. theoretically introduced multi-channel (MIMO) metasurface processors \cite{Abdolali2019, Babaee2021}. By engineering reciprocal bianisotropic responses, they demonstrated that a single metastructure could solve integro-differential equations and execute disparate mathematical tasks in parallel without crosstalk. The computational versatility of these non-local multilayer films recently culminated in the work of Liu et al. \cite{Liu2022}. Rather than being limited to a single mathematical operator, they engineered a single planar dielectric chip capable of performing multiple-order analog spatial differentiation as shown in Fig. 3(e). By meticulously tailoring the anisotropic angular transmission across different azimuthal directions in the momentum space, their multilayer chip can execute first-, second-, third-, and even fourth-order spatial derivatives simply by altering the polarization analyzer or illumination angle. This breakthrough proves that highly complex, multi-order mathematical operations can be intrinsically encoded into the non-locality of a single, ultra-thin unstructured film.

\subsubsection{\label{sec3.2.2}Resonant mechanisms}

While non-resonant interfaces offer robust, ultra-broadband operation, their angular dispersion is typically smooth and gradual, which fundamentally limits the NA and spatial resolution of the processed images. To synthesize highly abrupt mathematical kernels and push the limits of operational contrast, researchers have turned to spatially nonlocal resonant architectures. By coupling incident light into localized or guided resonances, the device's transmission or reflection coefficient undergoes sharp variations across narrow angular spectrums. Over the past decade, the physical realization of these nonlocal resonances has evolved from lossy metallic plasmons to high-index dielectric Mie resonances, ultimately reaching the extremes of $Q$-factor engineering via bound states in the continuum (BIC).

\begin{figure*}[htbp]
	\centering
	\includegraphics[width=\linewidth]{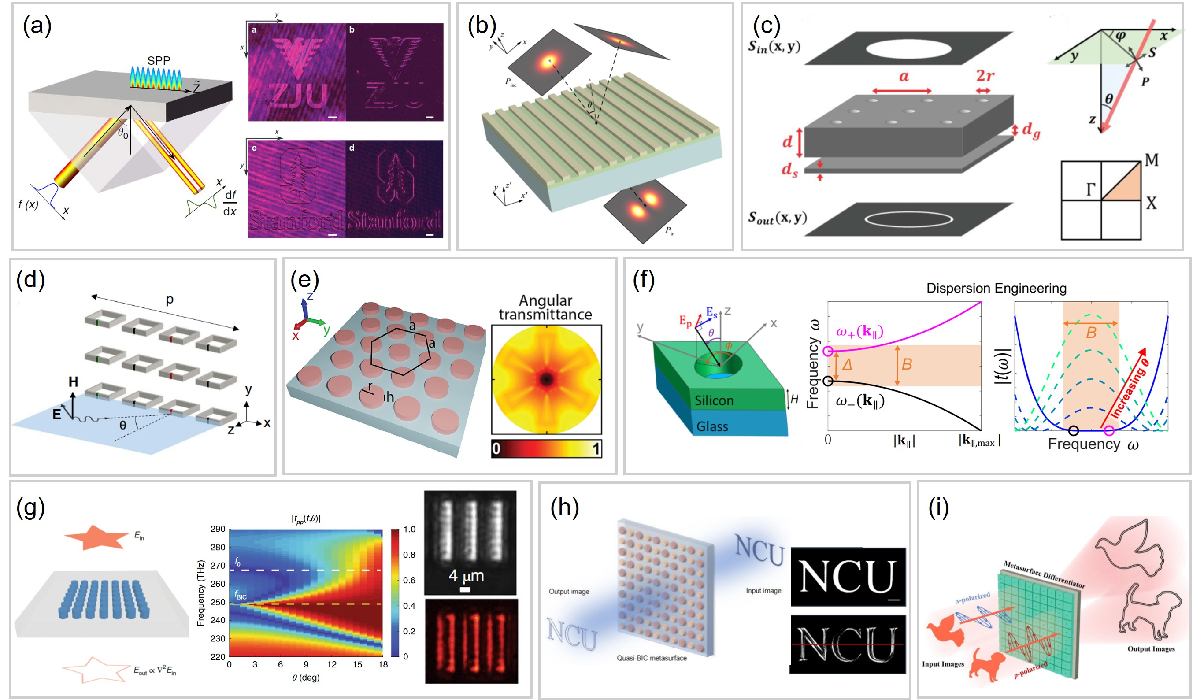}
	\caption{ Resonant metasurface platforms for nonlocal spatial-domain filtering. (a) An optical spatial differentiator based on SPPs \cite{Zhu2017}. (b) A spatial differentiator utilizing a guided-mode resonant diffraction grating\cite{Bykov2018}. (c) A 2D PhC slab operating as an isotropic Laplacian operator via guided resonances \cite{Guo2018}. (d) A high-index dielectric metasurface leveraging Fano resonances to manipulate signals in the momentum domain for mathematical operations \cite{Kwon2018}. (e) All-dielectric metasurfaces exploiting localized electric and magnetic Mie-type resonances for nanoscale edge detection \cite{Komar2021}. (f) A dispersion-engineered Laplacian metasurface achieving broadband, high-NA, highly efficient, isotropic, and dual-polarization edge detection \cite{Cotrufo2023a}. (g) A 2D photonic crystal slab executing a Laplacian operator by operating near a BIC \cite{Zhou2020}. (h) An all-dielectric metasurface harnessing quasi-BIC modes for high-quality, isotropic 2D spatial differentiation\cite{Liu2024a}. (i) A quasi-BIC-driven dielectric metasurface functioning as a dual-polarization Laplace differentiator \cite{Zhou2025a}.}
	\label{fig4}
\end{figure*}
 
The early exploration of resonant spatial domain filtering heavily relied on the excitation of surface plasmon polaritons (SPPs) in metallic structures. Because SPPs are highly sensitive to the in-plane wavevector of incident light, they naturally exhibit strong spatial nonlocality. Zhu et al. pioneered this approach by demonstrating a highly compact plasmonic metasurface capable of performing first-order spatial differentiation, as shown in Fig. 4(a) \cite{Zhu2017}. This concept was rapidly expanded across different spectrums and dimensions; for instance, Fang et al. utilized on-grating graphene surface plasmons to enable spatial differentiation in the terahertz region \cite{Fang2017}, while Hwang et al. designed a subwavelength plasmonic circuit for second-order spatial differentiation \cite{Hwang2018}. Concurrent work by Roberts et al. further highlighted the utility of metasurface dark modes in extracting specific spatial frequencies for image processing \cite{Roberts2018}.

While SPPs offer extreme subwavelength confinement and sharp nonlocality, they introduce fundamentally new physical constraints. Zhang et al. provided a critical theoretical insight into the time response of plasmonic spatial differentiators, revealing an intrinsic physical trade-off: the spatial resolution of the transfer function is fundamentally bounded by the temporal relaxation time of the SPP resonance or the $Q$-factor \cite{Zhang2019}. Despite these constraints, the mathematical versatility of resonant angular dispersion continues to expand. Lou et al., for instance, demonstrated that plasmonic nonlocality could be tailored beyond simple scalar derivatives to compute the divergence of vector fields, pushing analog computing into more complex vectorial mathematics \cite{Lou2020}. 

To mitigate intrinsic material losses and sharpen the angular response, the field transitioned to all-dielectric resonant platforms, initially utilizing periodic gratings and photonic crystals (PhCs). The foundational concept can be traced back to Bykov et al. (2014), who achieved the Laplace operator using a phase-shifted Bragg grating \cite{Bykov2014}. This was later advanced by Bykov who harnessed guided-mode resonances in dielectric gratings of Fig. 4(b) to perform highly efficient edge detection \cite{Bykov2018}. The similar approach was also demonstrated by Saba, and Fang \cite{Saba2018, Fang2018}.  Bezus et al. utilized high-Q resonances supported by dielectric ridges for precise beam differentiation \cite{Bezus2018}. Beyond 1D gratings, Fan's group extensively explored 2D photonic crystal slabs. For instance, Guo et al. demonstrated that by engineering band structure of the guided mode in a PhC slab in  Fig. 4(c), isotropic wavevector-domain image filters could be realized, perfectly emulating the Laplacian operator \cite{Guo2018}. Long et al. then expanded this to wavelength-controlled free-space compression and spatial differentiation by introducing a polarization-independent nonlocal metasurface by designing guided resonances with degenerate band curvatures in a photonic crystal slab \cite{Long2022}.

While PhC slabs rely on guided modes distributed over macroscopic areas, the drive for ultimate compactness led to the development of subwavelength dielectric metasurfaces governed by localized Mie resonances. Alù’s group significantly propelled this frontier by a series of works \cite{Kwon2018,Cordaro2019,Kwon2020,Cotrufo2023a,Cotrufo2023b}. Kwon et al. elegantly demonstrated that by designing an array of split-ring resonators in Fig. 4(d), optical nonlocality could be engineered at the nanoscale through the Fano-like interference\cite{Kwon2018}. This localized resonant engineering enables pristine mathematical operations. Such Fano resonant approach was then extended to high-index dielectric metasurfaces \cite{Cordaro2019,Kwon2020,Cotrufo2023a,Cotrufo2023b}. The resonant approach rapidly obtained attentions for high-fidelity spatial filtering. Zhou et al. utilized a high-quality magnetic resonance mode  hybridized with the classic bounded surface wave via grating coupling to implement first-order differentiation \cite{Zhou2019Analog}. Wan et al. exploited the spatial dispersion of electric dipole resonance supported by the silicon nanodisks in the metasurface, and realized optical analog computing of spatial differentiation and edge detection for two dimensions and arbitrary polarization at the visible wavelength \cite{Wan2020}. Further increasing the versatility of these structures, Kwon et al. designed dual-polarization nonlocal metasurfaces capable of parallel 2D image processing, proving that Mie-type resonances could simultaneously couple with polarization degrees of freedom \cite{Kwon2020}. Komar et al. utilized the strong angular dependence of the lowest electric and magnetic dipole resonances in silicon metasurfaces of Fig. 4(e) to perform highly transparent, transmission-mode edge detection \cite{Komar2021}. Cotrufo et al. presented an elegant way of dispersion engineering to design resonant metasurface  whose critical metrics are simultaneously optimized, as shown in Fig. 4(f)  \cite{Cotrufo2023a}. They also experimentally demonstrated silicon metasurfaces performing isotropic and dual-polarization edge detection, with high NA, broad spectral bandwidths, and large throughout efficiencies. Liu et al.realized polarization-independent for isotropic two-dimensional second-order differentiation by leveraging the magnetic dipole resonance \cite{Liu2025c}.

As the mathematical demands for transfer function engineering became more rigorous, requiring exact pole and zero alignments in the complex wavevector plane, researchers sought resonant states with theoretically infinite $Q$-factors. As a result, relentless innovation in modal engineering continues to push the boundary. Goh and Alù demonstrated that even a single nonlocal scatterer could act as a compact wave-based computer \cite{Goh2022}. Most recently, Zhou et al. achieved the Laplace differentiator by harnessing toroidal dipole resonances \cite{Zhou2024}. In particular, the concept of BICs has emerged as a powerful and highly customizable paradigm for nanophotonic optical computing. Because true BICs exist as embedded eigenstates completely decoupled from far-field radiation, introducing intentional structural perturbations transforms them into leaky quasi-BICs. This evolution imparts finite, highly controllable Q-factors and provides abundant degrees of freedom to accurately engineer both the spectral line width and the angular dispersion of the optical resonances.

The foundational proof-of-concept for this approach was established by Valentine’s group, who achieved a major breakthrough by employing a silicon metasurface operating near a symmetry-protected BIC shown in  Fig. 4(g) \cite{Zhou2020}. By slightly breaking the structural symmetry, they coupled incident light into a quasi-BIC Fano resonance, yielding an relatively sharp angular dispersion near the quasi-BIC resonance that functions as a precise spatial differentiator. Following this paradigm, Pan et al. formalized the application of quasi-BICs for Laplace metasurfaces \cite{Pan2021}. To fully exploit the inherent advantages of the BIC mechanism, our recent work proposed an all-dielectric metasurface that simultaneously engineers the Q-factor and the angular dispersion of the quasi-BIC resonance, as shown in Fig. 4(h) \cite{Liu2024a}. Composed of a highly symmetric square lattice of four silicon nanodisks per unit cell, the metasurface leverages specific structural perturbations via varying the radius of specific nanodisks to precisely tailor its angle-dependent transmission. This careful manipulation of the quasi-BIC's spatial dispersion yields a high-pass optical transfer function that perfectly matches the isotropic Laplacian operator. Consequently, we experimentally demonstrated high-quality, polarization-independent, and isotropic two-dimensional spatial differentiation without the need for any additional polarizing elements, paving a highly efficient and CMOS-compatible route for ultracompact optical image processing. The nonlocal metasurface based on quasi-BICs was also extended to realize broadband Laplace differentiator by Zhou et al., as shown in Fig. 4(i) \cite{Zhou2025a}. These advanced multipolar resonant states represent the current pinnacle of nonlocal flat optics, offering extreme sensitivity, near-zero loss, and unparalleled edge enhancement for the next generation of analog machine vision.

\subsection{\label{sec3.3} Interferometric difference platforms}

An alternative to direct frequency filtering in Fourier and spatial domains is the optical interferometric difference approach. Instead of selectively filtering spatial frequencies, this method captures the variation between discrete quantities. Mathematically, the spatial derivative $f'(x)$ can be closely approximated using the central finite difference method: $f'(x) \approx \frac{f(x+h)-f(x-h)}{2h}$, where $h$ signifies a tiny spatial displacement. To implement this optically, flat optics leverage multidimensional multiplexing to generate two slightly shifted copies of the original signal in different channels such as the orthogonal polarization states and coherently subtract them.

\begin{figure*}[htbp]
	\centering
	\includegraphics[width=\linewidth]{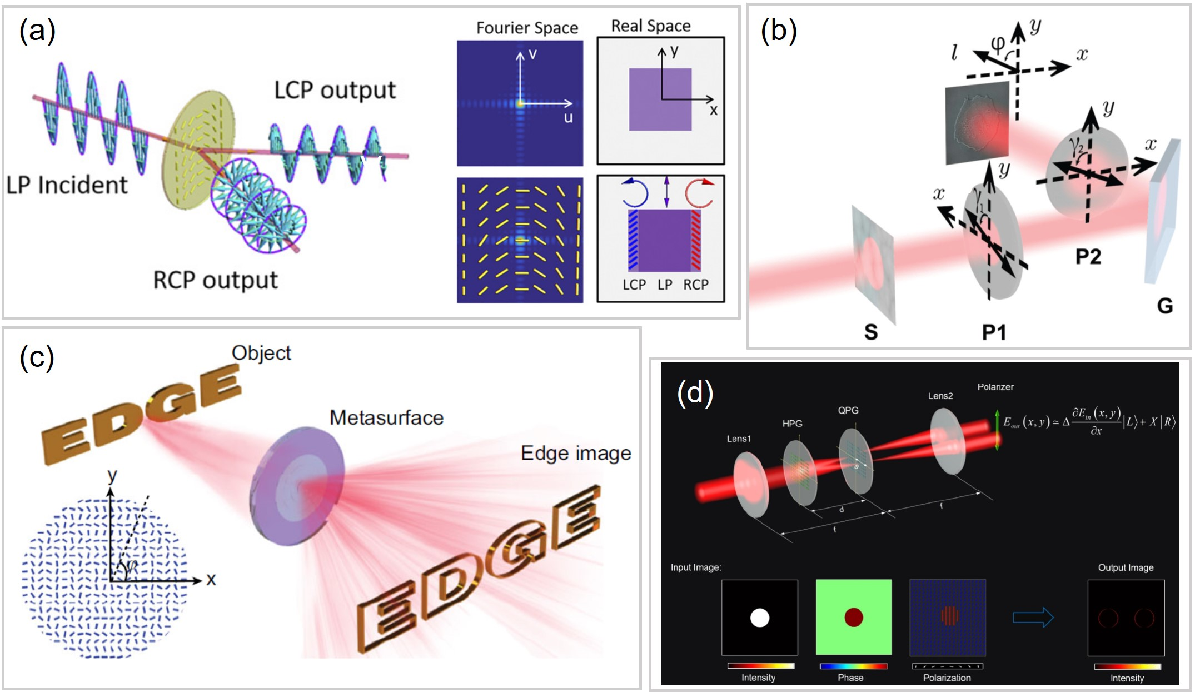}
	\caption{ Interferometric difference platforms leveraging spin-orbit interactions and phase manipulation. (a) A PB phase metasurface acting as an optical spatial differentiator by imparting opposite transverse shifts to orthogonal circular polarizations \cite{Zhou2019b}. (b) An actively adjustable spatial differentiator exploiting the photonic spin Hall effect upon reflection from a single planar dielectric interface \cite{Zhu2020}. (c)  \cite{Zhou2020a}. (d)  \cite{He2025}. (c) A two-dimensional isotropic optical spatial differentiator utilizing expanded geometric phase configurations \cite{Zhou2020a}. (d) A full-dimensional optical spatial differential imaging system utilizing giant and tiny spin Hall effects generated by liquid-crystal polarization gratings \cite{He2025}. }
	\label{fig5}
\end{figure*}

The primary physical mechanism driving this finite difference approach is the photonic spin Hall effect (SHE) and the manipulation of photonic spin-orbit interactions. Zhu et al. provided the fundamental theoretical framework, demonstrating that the intrinsic SHE of light upon reflection or transmission at optical interfaces naturally induces a generalized spatial differentiation \cite{Zhu2019}. To practically harness this for high-contrast imaging, Zhou et al. introduced a groundbreaking geometric phase (PB phase) gradient metasurface \cite{Zhou2019b}. By sandwiching this metasurface between two orthogonally oriented linear polarizers, the incident linearly polarized beam is split into left circularly polarized (LCP) and right circularly polarized (RCP) components. The phase gradient imparts opposite transverse shifts to the LCP and RCP beams, as shown in Fig. 5(a). The second polarizer coherently recombines and subtracts these two shifted fields, effectively executing a spatial derivative. Crucially, because the geometric phase is dictated purely by the spatial orientation of meta-atoms rather than material dispersion, this interferometric architecture inherently operates across an ultra-broadband spectrum, successfully achieving achromatic edge detection.

Following this paradigm, researchers rapidly sought to make these optical differentiators structurally integrated. Zhu et al. advanced the field by proposing adjustable spatial differentiators where the degree of spatial shifts and operational boundaries could be actively tuned via the photonic SHE, as shown in Fig. 5(b) \cite{Zhu2020}. The similar approach was also demonstrated by Mi et al. \cite{Mi2020}. To transcend the limitations of 1D directional edge detection, Zhou et al. further expanded the geometric phase configurations to realize 2D isotropic spatial differentiation, as shown in Fig. 5(c) \cite{Zhou2020a}. In parallel, Zhu et al. introduced topological optical differentiators, mapping the computational capabilities to the topological charge of the spatial spectrum \cite{Zhu2021}. Zong et al. designed a two-dimensional optical differentiator for broadband edge detection covering the visible band\cite{Zong2023}. Chen et al. then proposed a method for optical edge detection under ambient illumination in a natural environment scene, by employing a joint modulation technique that utilizes metasurface for differential operation and spatial light modulator as phase modulation \cite{Chen2025Optical}. He et al. achieved full-dimensional optical spatial differential imaging in amplitude, phase, and polarization dimensions \cite{He2025}. By creating giant and tiny spin-Hall effects using liquid-crystal half-wave and quarter-wave polarization gratings, as shown in Fig. 5(d),  they separated the input field  into two circular polarization basis vectors for independent differentiation operations, and obtained  controllable contrast in the differential images of the two circular polarization basis vectors. Meanwhile, the concept of spin-multiplexed differentiation has recently transcended patterned nanostructures. Yang et al. demonstrated that the intrinsic spin-orbit coupling within a natural thin uniaxial crystal can be leveraged for broadband achromatic spatial differentiation \cite{Yang2024}. By simply inserting a thin uniaxial crystal (e.g., YVO$_4$) under normal incidence, the intrinsic spin-to-orbit conversion naturally embeds an optical vortex into the image field, performing a second-order topological spatial differentiation. Crucially, because this intrinsic coupling is highly wavelength-independent, it enables isotropic and completely achromatic 2D edge detection under standard incoherent white LED illumination, offering an extraordinarily simple yet powerful alternative to complex nanofabricated metasurfaces.

Pushing the concept of polarization multiplexing toward advanced phase imaging, Wang et al. elegantly upgraded the interferometric difference mechanism to realize single-shot isotropic differential interference contrast (i-DIC) microscopy \cite{Wang2023}. Unlike conventional DIC systems that rely on bulky birefringent prisms to generate a rectilinear shear—yielding only a 1D anisotropic spatial difference—their designed metasurface introduces a rotationally symmetric radial shear between two orthogonal linear polarizations. By coherently recombining these radially shifted optical fields through a linear analyzer, the system directly executes a 2D isotropic finite difference. This compact architecture completely bypasses the need for mechanical rotation, effortlessly converting invisible phase gradients into high-contrast amplitude boundaries for the label-free observation of transparent biological specimens, such as breast cancer tissues.

The robustness and versatility of these spin-dependent architectures have propelled their integration into highly advanced application scenarios. Remarkably, because the geometric phase operations are fundamentally coherent and preserve quantum entanglement, this spin-multiplexing mechanism has successfully transitioned into the quantum optics regime. As pioneered by Zhou et al. and Yang et al., illuminating a dielectric metasurface with polarization-entangled photon pairs enables nonlocally triggered quantum edge detection\cite{Zhou2020b,Yang2025}. A detailed discussion of such quantum-assisted analog computing, along with its extraordinary noise-resilient properties, will be elaborated in Section 4.3.

\subsection{\label{sec3.4}Spiral-phase spatial filtering systems}

While the aforementioned interferometric platforms beautifully extract boundaries based on intensity variations, they are primarily tailored for amplitude objects. But in biomedical imaging, transparent biological specimens, like living cells, are classified as phase objects, which alter the phase delay of the transmitting wave without significantly absorbing light.

\begin{figure*}[htbp]
	\centering
	\includegraphics[width=\linewidth]{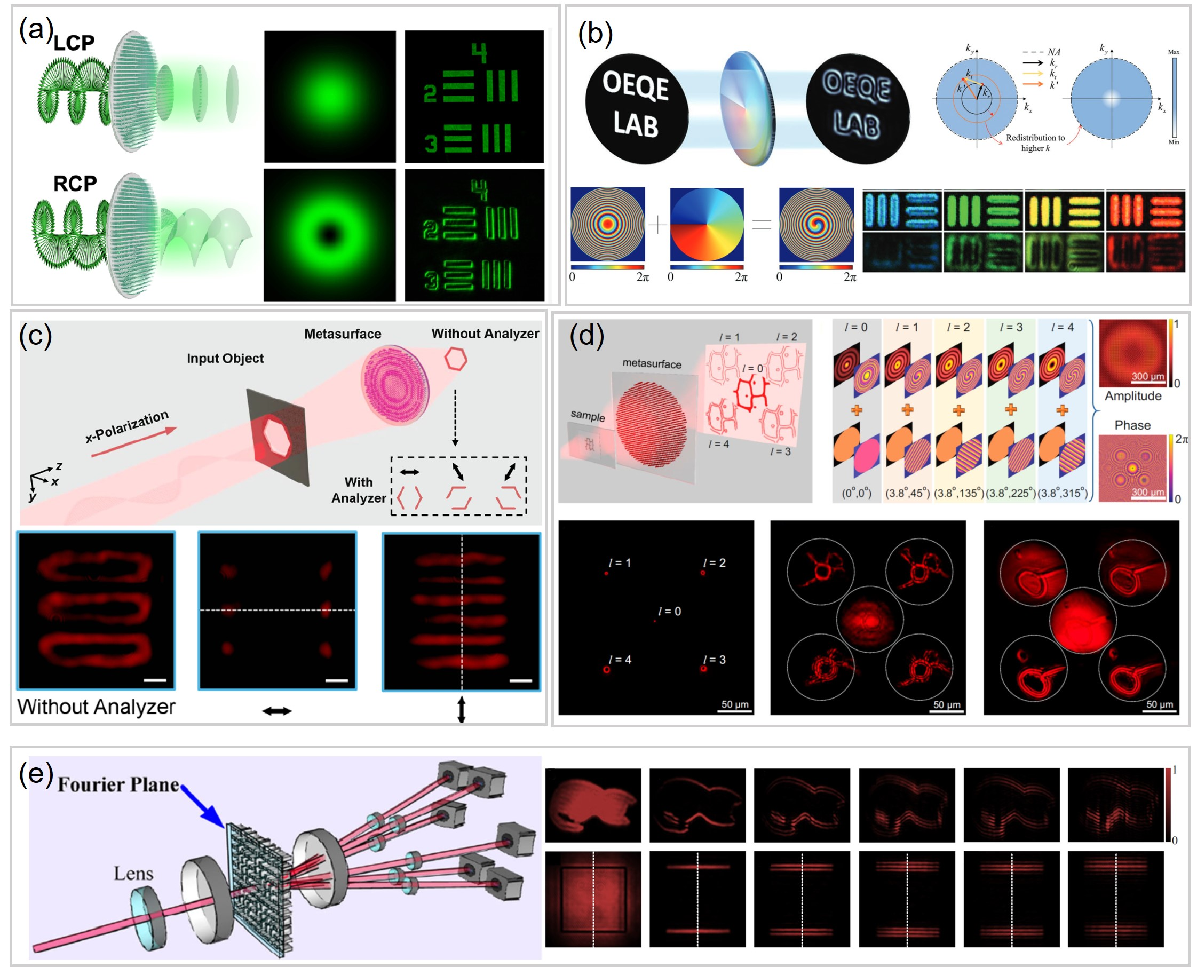}
	\caption{ Spiral-phase spatial filtering systems leveraging spiral phase contrast for analog computing.(a) A dielectric metasurface encoded with a spiral phase profile to execute isotropic 2D spatial differentiation \cite{Huo2020}. (b) A compact single-layer spiral metalens that superimposes a hyperbolic focusing phase with a topological charge-1 spiral phase to achieve direct edge-enhanced imaging \cite{Kim2022}. (c) A compact vector vortex metalens designed for real-time, broadband, direction-selective edge detection via the superposition of spin-dependent vortex and antivortex beams \cite{Ren2025}. (d) A Bessel vortex modulated metalens integrating complex amplitude modulation with Bessel vortex phase engineering for parallel multiple-order spatial differentiation \cite{Huo2024}. (e) A PB phase metasurface engineered to perform high-order optical spatial differentiation \cite{Qiu2025}.}
	\label{fig6}
\end{figure*}

The quest to visualize invisible phase gradients dates back to Zernike’s foundational work on phase contrast microscopy in 1942 and 1955 \cite{Zernike1942,Zernike1955}. To implement these massive classical interferometric setups within a compact analog computing framework, metasurfaces have been employed to engineer phase-sensitive transfer functions. The prominent technique of spiral phase contrast introduces an optical vortex into the Fourier plane. Huo et al. ingeniously integrated this with spin-orbit interactions, designing a photonic spin-multiplexing metasurface shown in Fig. 6(a) that can flexibly switch between bright-field imaging and isotropic spiral phase contrast imaging simply by altering the incident spin state \cite{Huo2020}. The designed metasurface can provide two uncorrelated phase profiles with a constant phase profile and a spiral phase profile corresponding to the two spin states of the incident light, and thus dynamically switchable ordinary diffraction imaging and isotropic edge- enhanced imaging are respectively demonstrated.

Targeting the bulky nature of conventional spatial light modulators and 4$f$ systems used in phase contrast imaging, Kim et al. proposed a compact single-layer spiral metalens in Fig. 6(b) \cite{Kim2022}. By elegantly superimposing a hyperbolic focusing phase with a topological charge 1 spiral phase, this single-layer metalens executes 2D isotropic edge-enhanced imaging directly. This approach not only completely removes the need for additional relay optics to access the Fourier plane but also achieves a submicrometer resolution of 0.78 $\mu$m over a broadband visible spectrum, representing a significant stride towards miniaturized analog optical processors. Following this compact design, Ren et al. presented a compact vector vortex metalens for real-time, broadband, direction-selective edge detection, by engineering the superposition of spin-dependent vortex and antivortex and beams as Fig. 6(c) shows\cite{Ren2025}.

Moving toward quantitative analysis, Ji et al. demonstrated quantitative phase contrast imaging by addition of a nonlocal angle-selective metasurface supporting guided mode resonance to a conventional microscop, overcoming the qualitative-only limitations of traditional edgeenhancement\cite{Ji2022}. As fabrication and inverse design advanced, these phase-sensitive architectures became highly integrated and precise. Fu et al. developed an ultracompact meta-imager capable of arbitrary all-optical convolution, enabling parallel spatial filtering and phase edge enhancement without separated cascaded lenses \cite{Fu2022}. The ultimate trajectory of this field is moving towards simultaneous, multi-domain information acquisition. Zhang et al. designed a multiplexed dielectric metasurface that synchronously performs spiral phase contrast and bright-field imaging in a single shot, providing perfectly registered amplitude and phase-edge details without moving parts \cite{Zhang2023}. Similarly addressing the demand for practical dual-mode observations, Sun et al. developed a near-infrared metalens-empowered microscope that realizes switchable spiral phase contrast and bright-field imaging simply via incident polarization control \cite{Sun2024}. Together, these phase-sensitive flat optics are redefining the boundaries of label-free, ultra-compact computational microscopy.

The massively parallel multiple-order computing was recently achieved using advanced meta-lens architectures. As shown in Fig. 6(d), Huo et al. pushed the boundaries of spatial multiplexing by integrating complex amplitude modulation with Bessel vortex phase engineering \cite{Huo2024}. By designing a single-layer metasurface that combines a focusing metalens with varying topological charges of Bessel vortices, they demonstrated a Bessel vortex modulated metalens. The order of the radial differentiation is elegantly determined by the topological charge of the encoded vortex. More importantly, they utilized angle multiplexing to create multiple independent information channels on a single chip, allowing zeroth- to fourth-order spatial differential operations to be executed synchronously and in parallel over a broadband visible spectrum. 

Beyond singular high-order operations, manipulating multiple arbitrary orders simultaneously remains a hallmark of advanced meta-processors. Qiu et al. successfully integrated multiple high-order differential operators onto a single chip as shown in Fig. 6(e)\cite{Qiu2025}. By synergizing PB phase with angle multiplexing via multi-beam interference phase patterning, their single metasurface synchronously executed first-, second-, and third-order spatial differentiations. The output images were cleanly separated into different diffraction angles, offering a highly scalable solution for massively parallel, multi-order image processing without the need to swap physical filters. This represents a paradigm shift toward massively parallel, ultra-fast computational kernels operating entirely at the speed of light.

\section{\label{sec4}Advanced meta-processors: parallel, dynamic and nonlinear computing}

As optical analog computing transitions from proof-of-concept demonstrations to practical system-level applications, the reliance on static, single-functional components has become a critical bottleneck. Modern machine vision, autonomous driving, and advanced biological imaging demand computational frameworks that can process multidimensional data streams concurrently or adapt their processing kernels in real-time. Consequently, the frontier of computational flat optics has shifted toward parallel multiplexing, dynamic reconfigurability, and nonlinear regime, leveraging advanced meta-architectures and active materials to manipulate the OTF on demand.

\subsection{\label{sec4.1} Parallel multiplexing computing}

Instead of relying on cascaded bulk optics or swapping optical filters to extract different features, parallel multiplexing encodes multiple independent mathematical operators into a single flat device. By exploiting various degrees of freedom of light, such as polarization, spin, and wavelength, these meta-processors can simultaneously execute distinct computational tasks without crosstalk.

\begin{figure*}[htbp]
	\centering
	\includegraphics[width=\linewidth]{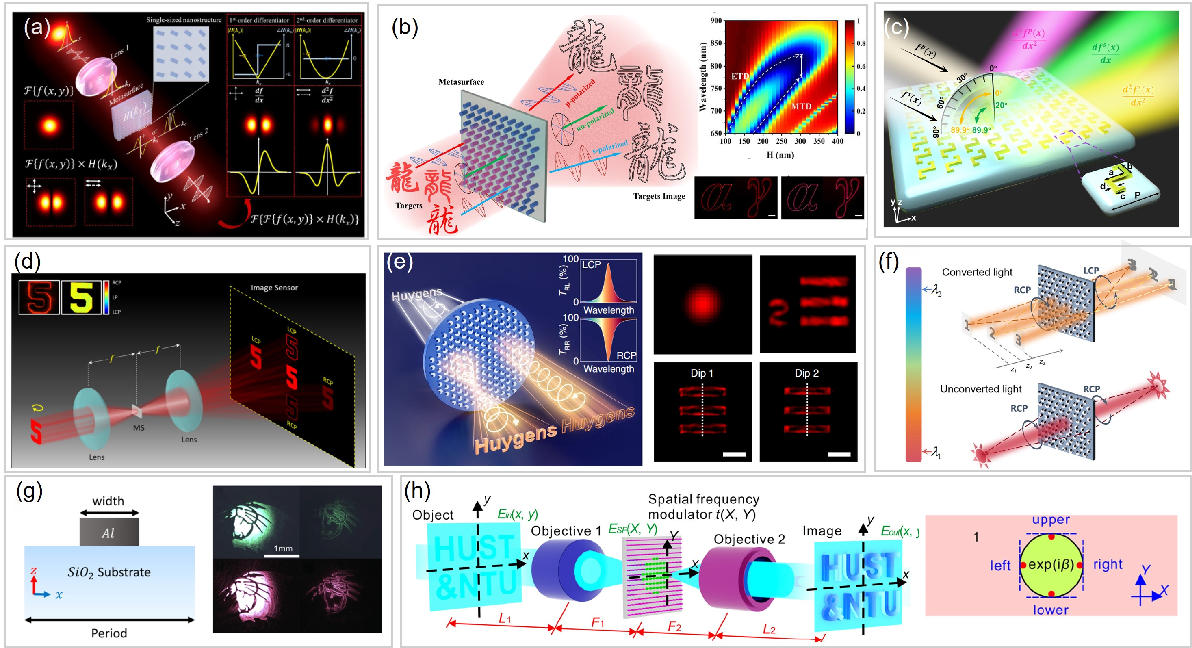}
	\caption{ Advanced meta-processors enabling parallel and multiplexed analog computing. (a) A multiplexed metasurface executing simultaneous outline detection (1st-order) and edge positioning (2nd-order) along orthogonal polarization axes \cite{Liang2023}. (b) A dual-polarization nonlocal metasurface based on Mie resonances for parallel 2D edge detection \cite{Zhou2025}. (c) A multifunctional ultra-wide-angle metasurface performing simultaneous first- and second-order spatial differentiation via polarization multiplexing \cite{Liu2026}. (d) A "3-in-1" geometric metasurface combining spatial multiplexing and polarization splitting to simultaneously project five distinct images with different optical properties in a single shot \cite{Intaravanne2023}. (e) A nonlocal Huygens' metasurface that utilizes spin-multiplexing to concurrently perform wavelength-selective bright-field imaging and edge detection \cite{Yao2025}. (f) A nonlocal Huygens metasurface utilizing spin-multiplexing for simultaneous edge enhancement and depth sensing \cite{Guo2025b}. (g) Concentric ring-based metasurfaces sustaining isotropic 2D edge detection across a broadband visible frequency range under both coherent and incoherent illumination \cite{Tanriover2023}. (h) Metasurface designed using half-side phase-contrast technique for simultaneous phase-contrast and relief-like imaging across the visible spectrum \cite{Deng2024a}.}
	\label{fig7}
\end{figure*}

By decoupling the phase and amplitude responses along orthogonal polarization axes, metasurfaces can execute different differential equations simultaneously. A minimalist yet highly effective strategy was proposed by Liang et al., who developed an all-optical multiplexed meta-differentiator using a single-sized Malus metasurface in Fig. 6(a) \cite{Liang2023}. Composed of identical silicon nanobricks with spatially engineered orientation angles, the device modulates the amplitude of spatial frequencies directly. It performs 1st-order spatial differentiation on the $y$-polarized component and 2nd-order differentiation on the $x$-polarized component. By simply rotating a bulk-optic analyzer, the system enables tri-mode surface morphology observation: bright-field imaging, rough outline detection, and high-precision edge positioning. To break the bandwidth limits of dual-polarization computing, Zhou et al. utilized a nonlocal hollow metasurface to excite quasi-BICs \cite{Zhou2025a}. The distinct angular dispersion capabilities of the magnetic dipole and quadrupole resonances allowed the device to act as a broadband Laplace differentiator for both $p$- and $s$-polarized incident light, maintaining high-resolution 2D edge detection over a remarkably broad spectral range of 165 nm. Then they experimentally demonstrated that azimuthal-insensitive Laplace differential operations and dual-polarization second-order two-dimensional edge detection within a broadband operating range by exciting and detuning the electric toroidal dipole and magnetic toroidal dipole resonances in Fig. 6(b)\cite{Zhou2025}. Extending parallel multiplexing into the terahertz regime, Liu et al. proposed a multifunctional ultra-wide-angle metasurface in Fig. 6(c) capable of executing simultaneous first- and second-order spatial differentiation \cite{Liu2026}. By harnessing a synergistic mechanism that combines critical coupling and near-far-field multi-wave superposition, they successfully suppressed the Brewster-angle effect and broke conventional NA limits. Operating under orthogonal polarization modes, the device achieves an extraordinary NA approaching unity of 0.966 experimentally, and a spatial resolution limit of 1.27$\lambda$, paving the way for ultra-high-resolution, non-paraxial terahertz imaging and communication. Pushing the concept of polarization multiplexing toward practical machine vision under ambient conditions, Wang et al. demonstrated a polarization-multiplexed metalens integrated with a commercial polarization camera\cite{Wang2024}. Instead of relying on coherent optical interference, this meta-processor encodes two distinct optical transfer functions, including a standard focusing profile for $x$-polarization and a spiral-phase-modulated profile for $y$-polarization. By simultaneously capturing both polarization channels in a single shot and performing a simple digital subtraction, the system achieves isotropic edge detection under completely incoherent sunlight illumination. This hybrid optoelectronic approach drastically reduces computational complexity while bypassing the strict coherent laser requirements of traditional interferometric systems.

Photonic spin-orbit interactions provide another robust multiplexing scheme, particularly for phase-object imaging. Intaravanne et al. ingeniously integrated spatial multiplexing and polarization splitting by designing a geometric metasurface that imparts both a spiral phase profile and off-axis phase gradients \cite{Intaravanne2023}. Placed in the Fourier plane, this single device enables a "3-in-1" microscopy platform that simultaneously projects five spatially separated images in a single shot: one bright-field image, two edge-enhanced images with opposite circular polarizations via spiral phase contrast, and two conventional polarization images, as shown in Fig. 6(d). Similarly, Zhang et al. demonstrated a dielectric metasurface capable of synchronously performing spiral phase contrast and bright-field imaging within the same field of view, providing perfectly registered amplitude and phase-edge details in Fig. 6(e) \cite{Zhang2023}.  Sun et al. presented a compact dual-mode microscopes based on the metalens in the NIR imaging window,  which can be tuned between the spiral phase contrast imaging and bright field imaging by polarization control \cite{Sun2024}. 

To achieve high-quality and crosstalk-free multiplexed optical computing, nonlocal resonant mechanisms have recently been introduced. Yao et al. proposed a nonlocal Huygens' metasurface that utilizes spin-multiplexing to concurrently perform wavelength-selective bright-field imaging and edge detection \cite{Yao2025}, as shown in Fig. 7(e). By leveraging quasi-BICs and the generalized Kerker condition, the meta-lens effectively separates the output into a converted state with a focusing phase profile and an unconverted state acting as an angular-dispersion filter, overcoming traditional bandwidth and crosstalk limitations. Following this concept, our group attempted to perceive complex 3D environments by proposing a dynamically tunable nonlocal metasurface for synchronous depth sensing and edge enhancement in a single-shot exposure, as shown in Fig. 7(f) \cite{Guo2025}. Through spin-multiplexing, the converted cross-polarized light undergoes PB phase modulation to create a double-helix point spread function, quantitatively mapping the object's depth to the rotation angle of focal spots. Simultaneously, the unconverted light utilizes the metasurface's angular dispersion for momentum-space high-pass filtering, yielding a high-contrast edge image. Most recently, Yu et al. employed double-phase encoding and polarization multiplexing to enables arbitrary image transformations within a single passive nanophotonic device,including first-order differentiation, cross-correlation, vertex detection, and Laplacian differentiation \cite{Yu2026}.

Beyond polarization and spin, the illumination wavelength serves as another critical degree of freedom for parallel computing. However, resonant metasurfaces typically suffer from severe chromatic dispersion, restricting operations to narrow spectral bands. Overcoming this limitation enables ultra-broadband parallel processing, where multi-wavelength information is computed simultaneously without chromatic distortion. For instance, Tanriover et al. demonstrated that concentric ring-based metasurfaces in Fig. 6(g) could sustain isotropic 2D edge detection across the entire visible frequency range under both coherent and incoherent broadband illumination \cite{Tanriover2023}. Deng et al. recently proposed a dielectric metasurface capable of simultaneous phase-contrast and three-dimensional relief-like imaging across the visible spectrum shown in Fig. 6(h) \cite{Deng2024a}. By imparting an asymmetrical phase shift to the diffracted waves, their half-side phase-contrast  method seamlessly converts minute phase discrepancies into shadow-cast amplitude features, achieving high-quality multi-wavelength observation of transparent biological cells. Applying this ultra-broadband capability to entirely new mathematical domains, the same team experimentally realized the angular spectrum meta-processors conceptually introduced in Section 2.3.3 \cite{Deng2024}. By tailoring the anisotropic transmission coefficients of silicon nanopillars, they achieved first-order, second-order, and mixed-order partial derivatives directly in the momentum space on the angular spectrum of an image across the entire visible spectrum from 450 nm to 685.5 nm. This broadband angular filtering allows for the parallel isolation and enhancement of extremely weak spectral features from strong overlapping backgrounds. Recently, Bi et al. proposed an optical front-end processor integrating two coherent transfer functions corresponding to differential and integral convolution kernels into a built-in metasurface by polarization encoding, allowing concurrent processing of multiple all-optical computational tasks at multiple visible wavelengths \cite{Bi2025}. Together, these advancements highlight the immense potential of wavelength-independent meta-processors for full-color machine vision and massive parallel derivative spectroscopy.

\subsection{\label{sec4.2} Dynamically reconfigurable computing}

The ultimate goal of optical analog computing is to realize fully programmable optical processors analogous to electronic FPGAs. This requires dynamically altering the device's OTF post-fabrication. Recent breakthroughs have achieved this by integrating active materials and novel structural tuning mechanisms into computational metasurfaces.

\begin{figure*}[htbp]
	\centering
	\includegraphics[width=\linewidth]{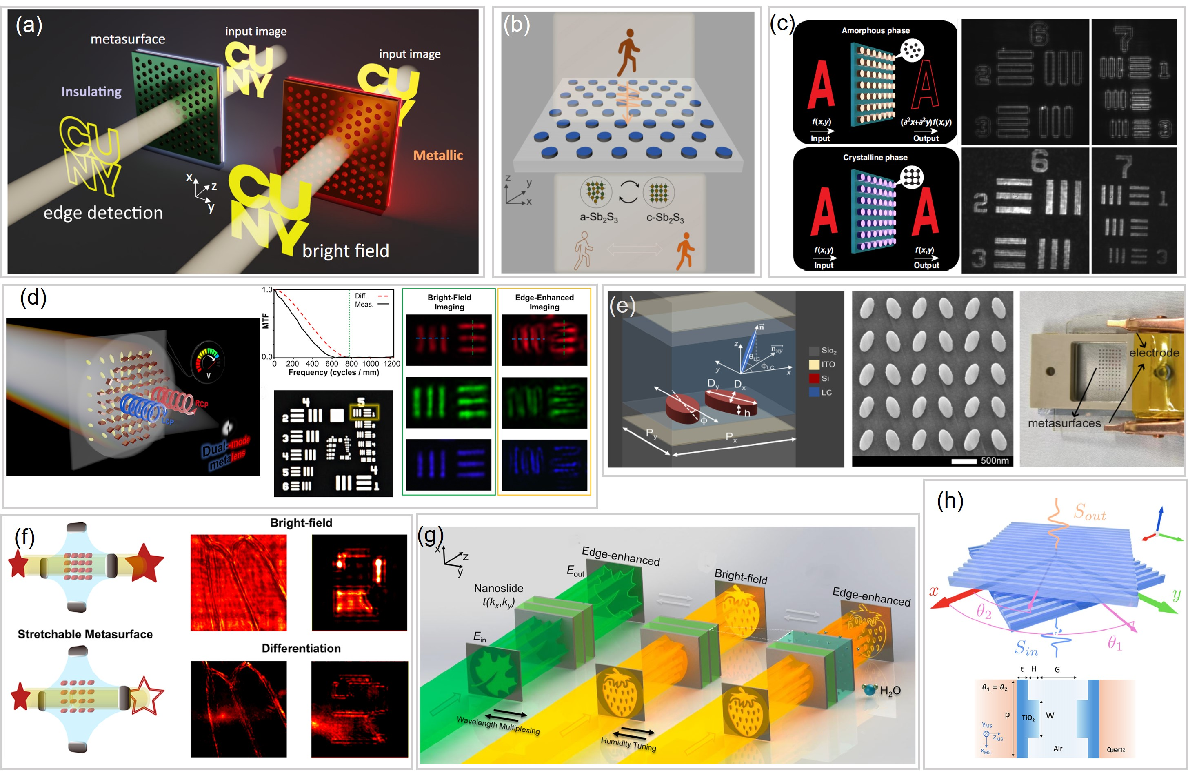}
	\caption{ Dynamically reconfigurable meta-processors for switchable analog computing. (a)–(c) Nonlocal metasurfaces incorporating phase-change materials, specifically (a) VO$_2$ \cite{Cotrufo2024}, (b) Sb$_2$S$_3$ \cite{Liu2025b}, and (c) Sb$_2$Se$_3$ \cite{Yang2025b}, to reversibly switch the optical transfer function between isotropic edge detection and conventional bright-field imaging via structural phase transitions. (d)–(e) A dual-mode metalens integrated with an electrically driven LC cell, enabling continuous switching between bright-field and high-contrast edge-enhanced imaging \cite{Badloe2023}\cite{Zhang2025}. (f) A mechanically stretchable metasurface embedded in an elastomer substrate, transitioning from a transparent imager to a 2D Laplacian differentiator by applying external strain \cite{Zhang2021}. (g) A hydrogel-scalable nanoslide that dynamically switches spatial-frequency processing regimes in response to variations in ambient humidity and illumination wavelength \cite{Dai2023}. (h) A moiré meta-optics platform comprising twisted bilayer metasurfaces, where adjusting the mutual twist angle continuously reconfigures the device's mathematical computation capabilities \cite{Wong2026}.}
	\label{fig8}
\end{figure*}

PCMs offer a non-volatile and highly robust mechanism for reconfiguring the spatial dispersion of metasurfaces. In the early research, Yang et al. theoretically demonstrated phase-change metasurfaces with a 4$f$  system for dynamic switchable 2D edge-enhanced imaging and bright-field imaging through selecting two independent phase profiles in amorphous and crystalline states \cite{Yang2021}. As shown in Fig. 7(a), Cotrufo et al. demonstrated a reconfigurable image processor by embedding a thin layer of vanadium dioxide (VO$_2$) within a silicon photonic crystal \cite{Cotrufo2024}. Below the transition temperature ($<$60$^\circ$C  ), the insulating VO$_2$ allows long-range interactions between unit cells, supporting nonlocal guided resonances that act as a high-pass filter for isotropic edge detection. Upon heating above 68$^\circ$C, the VO$_2$ transitions to a metallic phase; the increased optical loss inhibits nonlocality, instantly flattening the angle-dependent transmission profile to restore unfiltered bright-field imaging. To transition these capabilities from the infrared to the visible spectrum, where VO$_2$ and GST suffer from high losses, recent works have adopted antimony sulfide (Sb$_2$S$_3$). Our group then demonstrated Sb$_2$S$_3$-based nonlocal metasurfaces that operate entirely in real-space without 4$f$ configurations \cite{Liu2025b, Guo2025b}. The amorphous state of Sb$_2$S$_3$ exhibits strong angular dispersion driven by magnetic and electric Mie-type resonances, functioning as a Laplacian filter for efficient edge detection. Upon transitioning to the crystalline state, the nonlocality is reduced, yielding an angle-independent profile for uniform bright-field imaging. This reversible phase transition enables dynamic switching of the computational kernel in the visible regime, as shown in Fig. 7(b). Chamoli et al. then shifted the OTFs of a nonlocal metasurface hoseted by Sb$_2$S$_3$ to realize three imaging modalities \cite{Chamoli2025}. In our later work, to overcome the inherent narrowband limits of resonant flat optics, integrating the phase-change material Sb$_2$S$_3$ enables dynamic tuning of the operational wavelength across a 100 nm bandwidth \cite{Guo2025}. Further enriching the library of active materials, Yang et al. and Kendall et al. demonstrated a nonvolatile, reconfigurable nonlocal metasurface utilizing the low-loss phase-change material Sb$_2$Se$_3$ \cite{Yang2025b,Kendall2025}. By switching the Sb$_2$Se$_3$ nanopillars between their amorphous and crystalline states, the metasurface's optical transfer function is radically altered, from a Laplacian filter for isotropic edge detection to  a  flat transfer function for conventional bright-field imaging as shown in Fig. 7(c). Operating in the near-infrared region, this dual-functionality platform provides a high numerical aperture ($\sim$0.5) and retains its state without continuous energy consumption, marking a significant step towards all-solid-state, zero-static-power programmable meta-optics.

For high-speed, continuous, and damage-free tuning, electro-optic platforms utilizing liquid crystals (LCs) and graphene are highly desirable. Breaking away from complex nanostructured metasurfaces, Badloe et al. proposed an elegant, all-LC computing platform \cite{Badloe2023}. By modulating the external AC voltage applied to the LC phase plate, the phase profiles can be switched between the functions of bright-field and high-contrast edge-enhanced imaging, as shown in Fig. 7(d). The simliar concept also applied to the interference control by Liu et al. \cite{Liu2024}. The designed system pairs a static LC polarization grating, which induces spin-dependent spatial separations via the PB phase, with an electrically switched LC phase plate. By modulating the external AC voltage, the phase retardance is precisely tuned on a microsecond scale. This dynamically alters the interference conditions of the split beams, allowing the user to continuously and smoothly switch the output from a bright-field image to a high-contrast 1D or 2D edge-enhanced image. Integrating LCs directly with resonant metasurfaces provides profound control over the momentum-space dispersion. Recently, Zhang et al. experimentally demonstrated an electrically tunable, LC-integrated quasi-BIC metasurface in Fig. 7(e) operating in the near-infrared \cite{Zhang2025}. By applying an external voltage, the orientation of the nematic LC directors is modulated, which actively tailors the Q-factor and angular dispersion of the nonlocal quasi-BIC modes. This CMOS-compatible metadevice achieves fast-responsive millisecond-scale switching between standard bright-field imaging and high-contrast 1D edge detection, marking a significant step toward practical, high-speed reconfigurable optical computing.  Liu et al. utilized planar liquid crystal Alvarez lens to  enable adjustable resolution edge detection with an edge width from 32.5 $\mu$m to 73.7 $\mu$m without requiring any rotation or axial displacement\cite{Liu2025a}. Beside LC materials, graphene has also been  widely used for reconfigurable image processing \cite{AbdollahRamezani2015,ZangenehNejad2017,Xia2023}. At the nanoscale, Xia et al. theoretically demonstrated that gating a monolayer of graphene on a quartz substrate can tune its Fermi energy, actively modulating the complex Fresnel reflection coefficients near the Brewster angle \cite{Xia2023}. This subtle modulation actively controls the photonic spin Hall effect, dynamically switching the spatial differentiation output from 1D directional edges to 2D isotropic contours.

Mechanical manipulation offers a direct route to continuously tune computing responses without altering intrinsic material properties. A pioneering demonstration of this approach was presented by Zhang et al., who developed a highly stretchable metasurface composed of silicon nanoposts embedded in a polydimethylsiloxane (PDMS) elastomer as shown in Fig. 7(f)\cite{Zhang2021}. By mechanically straining the substrate, the lattice period is dynamically altered, modifying the interparticle coupling and in situ switching the device's transfer function. This enables the device to smoothly transition from a highly transparent bright-field imager  to a 2D Laplacian differentiator for isotropic edge detection. Recently, a groundbreaking paradigm was introduced by Wong and Cotrufo , who utilized twisted bilayer metasurfaces in moiré meta-optics \cite{Wong2026}. By simply twisting two identical resonant waveguide gratings around a common axis, the coupling between the layers is radically altered, as shown in Fig. 7(h). The authors numerically proved that adjusting this mutual twist angle continuously reconfigures the temporal transfer function, seamlessly switching the device's operation on an incoming optical pulse among zeroth-order, first-order, and second-order temporal differentiation.

Beyond mechanical rotation, environmental responsiveness provides a unique tuning mechanism. Dai et al. developed a hydrogel-scalable nanoslide comprised of alternating silver and polyvinyl alcohol (PVA) hydrogel layers shown in Fig. 7(g)\cite{Dai2023}. Operating directly in the wavevector domain, this macroscopic thin-film stack acts as an angular-dependent Fabry-Perot cavity. As ambient humidity increases, the hydrogel layers absorb water and swell, physically expanding the cavity thickness. This swelling predictably shifts the angular transmittance spectra, effectively tuning the spatial-frequency filter from a low-pass regime for bright-field imaging to a high-pass regime for edge detection, offering a highly practical, sensor-integrated approach to dynamic image processing.

\subsection{\label{sec4.3} Nonlinear analog computing}

Conventional analog computing metasurfaces operate almost exclusively in the linear optical regime, meaning the input and output optical frequencies are identical, and the spatial transfer function is strictly independent of the incident light intensity. To break this fundamental limitation, nonlinear computational flat optics has emerged. It not only introduces novel contrast mechanisms and wavelength conversion capabilities, e.g., infrared-to-visible up-conversion, but also enables intensity-dependent dynamic tunability. Over the past few years, nonlinear optical analog computing has experienced a clear evolutionary transitioning from bulky traditional crystal setups to unpatterned films, and ultimately advancing towards ultra-thin, multifunctional integrated metasurface architectures.

\begin{figure*}[htbp]
	\centering
	\includegraphics[width=\linewidth]{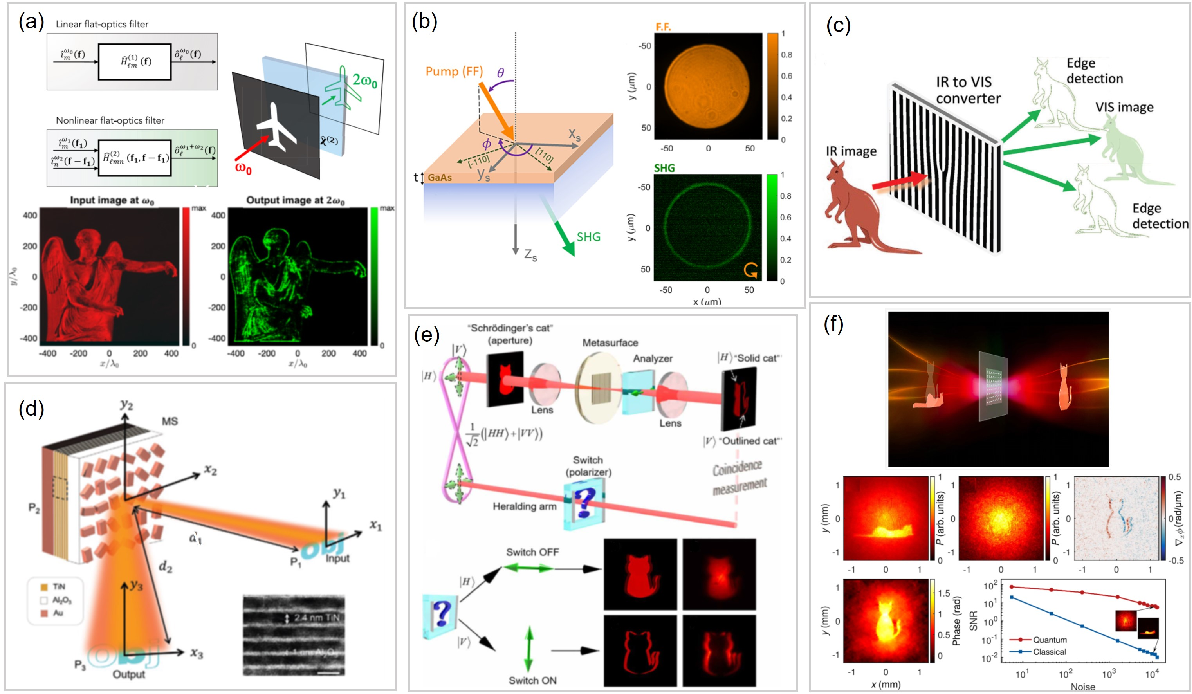}
	\caption{ Advanced meta-processors for nonlinear and quantum-assisted analog computing. (a) A nonlinear flat-optics platform performing edge detection at the second-harmonic frequency by synthesizing Volterra kernels for spatial filtering \cite{Ceglia2023}. (b) An unpatterned, nonlinear GaAs thin film bonded to a quartz substrate, leveraging intrinsic nonlocality for isotropic edge enhancement \cite{Cotrufo2025}. (c) Schematic of a monolithic nonlocal metasurface utilizing sum-frequency generation (SFG) and a phase dislocation to simultaneously output a direct image (zeroth order) and a differentiated edge image ($\pm$1 diffraction orders) \cite{Molina2024}. (d) An intensity-dependent nonlinear computational metalens, which outputs a high-contrast edge image at low intensities and transitions to a diffracted bright-field image at high intensities \cite{Zhou2022}. (e) A quantum-assisted optical edge detection system, where a dielectric metasurface illuminated by polarization-entangled photons acts as a nonlocal quantum switch to conditionally trigger the edge detection mode \cite{Zhou2020b}. (f) Quantum phase image distillation utilizing an integrated computing metasurface with symmetry-breaking nanostructures. By solving the 1D Poisson equation through quantum entanglement, overwhelming amplitude noise (dead cat) is removed to rapidly distill a high-contrast pure phase signal (alive cat) \cite{Yang2025}.}
	\label{fig9}
\end{figure*}

The concept of nonlinear spatial filtering was initially validated in traditional bulk nonlinear crystals. In 2018, Qiu et al. laid the conceptual foundation for nonlinear phase-contrast imaging \cite{Qiu2018}. By introducing a spiral phase plate into the pump beam path of a 4$f$ system and utilizing harmonic generation within a potassium titanyl phosphate (KTP) crystal, they demonstrated that invisible infrared illumination could be used to observe pure phase objects, detecting their edges in the visible spectrum. Subsequently, Liu et al. combined spiral phase contrast  operations with sum-frequency generation (SFG) to achieve field-of-view (FOV) enhanced infrared up-conversion edge detection in a periodically poled KTP (PPKTP) crystal \cite{Liu2019}. Although these pioneering works successfully proved the feasibility of nonlinear optical image processing, they relied heavily on bulky 4$f$ cascaded systems consisting of multiple lenses, external phase plates, and thick crystals, which fundamentally hindered their integration into compact machine vision systems.

To eliminate the reliance on bulky optical components, researchers began to exploit the intrinsic nonlocality of the materials themselves. De Ceglia et al. established a critical theoretical framework in Fig. 8(a), demonstrating that combining nonlinear phenomena with engineered nonlocality in flat optics allows for the synthesis of Volterra kernels that transcend linear Fourier optics \cite{Ceglia2023}. Guided by this framework, Cotrufo et al. experimentally demonstrated that even unpatterned anisotropic nonlinear films can perform analog computing as shown in Fig. 8(b) \cite{Cotrufo2025}. By leveraging the intrinsic nonlocality of the $\chi^{(2)}$ tensor in unpatterned gallium arsenide (GaAs) films during second-harmonic generation (SHG), they achieved broadband, polarization-selective nonlinear edge detection of input images directly on the untreated film, without the need for any nanoscale structural etching. This approach gracefully overcomes the narrowband limitations inherent to the high-Q resonances of linear metasurfaces, enabling nonlinear computing over an exceptionally broad spectral range.

To simultaneously achieve extremely high nonlinear conversion efficiency and complex transfer function control at the nanoscale, resonant dielectric metasurfaces have emerged as the ideal platform. Molina et al. achieved a milestone breakthrough by designing a nonlocal metasurface up-converter based on a high-$Q$ lithium niobate (LiNbO$_3$) thin film \cite{Molina2024}. By ingeniously utilizing the coherent nature of SFG in conjunction with the high-Q guided-mode resonances of the metasurface, they attained a record-high infrared-to-visible conversion efficiency but also empowered the metasurface with spatial multiplexing capabilities through the introduction of a one-dimensional topological dislocation. This monolithic device would directly output the bright-field up-converted image of an infrared object in the zeroth-order diffraction channel, while synchronously generating the differentiated edge-detection image in the $\pm$1 diffraction orders, as shown in Fig. 8(c). This concept signifies the true entry of nonlinear optical computing into the era of ultra-compact monolithic metasurface integration.

Beyond frequency conversion, another core advantage of nonlinear metasurfaces lies in the dynamic evolution of the transfer function in response to incident light intensity. Zhou et al. experimentally realized a computational edge-detection metalens based on third-order nonlinearity of the Kerr effect \cite{Zhou2022}. The metalens consists of gold nanoantennas featuring geometric phases stacked with metallic multiple quantum wells shown in Fig. 8(d). The quantum well layer provides a profound, intensity-dependent dynamic phase shift. At low illumination intensities, the coherent transfer function (CTF) of the device acts as a high-pass filter, outputting high-contrast edge images. As the incident intensity increases, the Kerr effect alters the refractive index of the material, and the CTF smoothly transitions into an all-pass filter, seamlessly switching the output to a conventional diffracted bright-field image of the object. This characteristic of dynamically altering the computational kernel simply by adjusting the laser power offers a groundbreaking paradigm for fully programmable optical computing.

Beyond classical nonlinear frequency conversion and intensity-dependent tuning, the frontier of advanced meta-processors is rapidly extending into the realm of quantum optics. The integration of analog computing metasurfaces with entangled photon sources offers unprecedented capabilities for ultra-weak light imaging and noise-resilient processing. As briefly introduced in Section 3.3.1, Zhou et al. pioneered this transition by demonstrating quantum-assisted edge detection in Fig. 8(e) \cite{Zhou2020b}. By illuminating a spin-multiplexed dielectric metasurface with polarization-entangled photon pairs, they realized a nonlocal quantum switch for spatial differentiation. Merely by selecting the polarization state of $|H⟩$ or $|V$⟩ of the idler photons in a remote heralding arm, the imaging system conditionally triggers either a normal bright-field image or an edge-enhanced image of the target. This architecture not only establishes a novel mechanism for secure image communication but also delivers a remarkably high signal-to-noise ratio (SNR) by eliminating uncorrelated background noise through temporal coincidence measurements, making it highly desirable for photon-hungry biological imaging. Building upon this quantum foundation, recent advancements have pushed towards more complex phase-sensitive quantum operations. Yang et al. integrated computational metasurfaces with spontaneous parametric down-conversion (SPDC) to propose a propose a quantum phase distillation scheme in Fig. 8(f)\cite{Yang2025}. While traditional quantum spatial correlation computing requires extensive integration times, their approach utilizes an integrated computing metasurface to directly solve the 1D Poisson equation as the photons traverse it. Combined with the coincidence imaging of entangled photons, this method not only extracts the gradient of a pure phase signal extremely rapidly but also exhibits remarkable noise robustness. Even when the noise intensity is two orders of magnitude higher than the signal, it successfully distills high-contrast phase images. This work not only highlights the immense potential of metasurfaces as analog computing cores in quantum information processing but also opens a new frontier for label-free quantitative phase imaging of biological cells.

\section{\label{sec5}Conclusions and outlooks}

In conclusion, metasurface-empowered optical analog computing has experienced a spectacular evolution, emerging as a disruptive solution to the latency and energy bottlenecks of conventional electronic computing. This Review has systematically distilled the developmental trajectory of this field, tracing its roots from fundamental transfer function engineering to the realization of highly advanced, multidimensional meta-processors. We have witnessed a profound architectural paradigm shift from macroscopic, lens-dependent Fourier-domain filtering to ultra-compact, real-space operations utilizing nonlocal spatial domain dispersion and spin-dependent interferometric architectures.

More importantly, the scope of optical analog computing has transcended simple 1D or 2D edge extraction of amplitude images. Driven by the relentless demand for high-throughput machine vision, the hardware itself has evolved from static, linear, and single-task filters into massively parallel, dynamically reconfigurable, nonlinear, and quantum-assisted computing platforms. Despite these tremendous successes in theoretical frameworks and proof-of-concept demonstrations, transitioning from laboratory prototypes to large-scale, real-world machine vision systems still faces formidable physical and engineering hurdles. Moving forward, the evolutionary trajectory of this field will inevitably gravitate towards the following strategic directions.

\subsection{\label{sec5.1}Physical mechanisms and inverse design for optimal transfer functions}

Historically, the physical mechanisms of flat optics have faced a stringent trade-off between the NA, spatial resolution, and operational bandwidth. To break these limitations, researchers are actively seeking novel resonant mechanisms to tailor the OTF. A highly promising frontier is the utilization of topological optical states, such as quasi-BICs. By judiciously tuning the structural asymmetry of meta-atoms, one can precisely engineer the resonant $Q$-factor and the sharpness of the angular dispersion. This unprecedented control over spatial nonlocality allows for the exact customization of the computing resolution and NA, enabling ultra-sharp edge detection that was previously unattainable.

At the same time, the design paradigm is shifting from intuition-based forward design to AI-assisted inverse design. Relying on brute-force parameter sweeps is no longer sufficient to approach theoretical limits. By leveraging data-driven algorithms, such as adjoint optimization, topology optimization, and deep generative models\cite{Chu2023,Liu2024b,Swartz2024}, researchers can explore non-convex parameter spaces to uncover freeform nanostructures. This AI-driven approach not only facilitates the precise synthesis of complex, higher-order mathematical kernels but also finds the global optima to balance large aperture, high throughput efficiency, and broadband operation.

\subsection{\label{sec5.2} Degrees of freedom expansion for multidimensionality and fully programmable optical computing }

The ultimate computing capacity of a metasurface is fundamentally bottlenecked by its degrees of freedom (DOFs). The inevitable trend for next-generation analog processors is the transition toward fully programmable, nonlinear, and multifunctional platforms. First, true complex computing tasks like deep learning classification, heavily rely on nonlinearity. Transitioning OTFs from the linear regime to the nonlinear and quantum regimes by means of incorporating intensity-dependent signal amplification, wavelength conversion, and entangled-photon noise resilience, is crucial for processing dynamic and ultra-weak-signal targets. Second, the inability to alter the computational kernel post-fabrication currently limits device versatility. By integrating active materials, such as phase-change materials like GST/VO$_2$/Sb$_2$S$_3$, electro-optic polymers, or nematic liquid crystals, the spatial and temporal weights of the metasurface can be continuously and electrically tuned on demand. This paves the way for an "optical FPGA". Furthermore, utilizing spatiotemporal, polarization, and ultra-broadband wavelength multiplexing to execute disparate mathematical tasks in parallel without crosstalk will maximize the information density of a single ultra-thin chip, evolving single-function filters into massively parallel meta-imagers.

\subsection{\label{sec5.3} System-level integration with optical neural networks}

As flat optics transition from laboratory prototypes to real-world machine vision, hybrid optoelectronic architectures are emerging as a practical milestone. For instance, Wang et al. recently integrated a polarization-multiplexed metalens directly with a commercial polarization camera to serve as an analog front-end for edge detection under ambient sunlight \cite{Wang2024}. By offloading the complex spatial filtering to the metalens and leaving only a low-cost point-by-point subtraction to the digital backend, such optoelectronic processors perfectly illustrate how meta-optics can alleviate the computational burden on subsequent digital systems in real-time target detection tasks.

Optical analog processors are perfectly poised to serve as the zero-power, zero-latency front-end physical layer for ONNs. By performing instantaneous feature extraction such as edge/contour highlighting, phase-gradient extraction, and angular spectrum isolation, before the signal undergoes analog-to-digital conversion, metasurfaces aggressively filter out redundant background data, drastically alleviating the computational burden on backend electronic processors. Furthermore, these analog computing layers can be embedded into all-optical diffractive deep neural networks (D$^{2}$NNs) \cite{Qiu2026,Liang2026}. Acting as specialized convolutional or differential hidden layers, they can significantly reduce the required depth and complexity of the neural network while maintaining or even enhancing image recognition accuracy. Ultimately, synergizing customized meta-differentiators with advanced optical network architectures will unlock unprecedented capabilities for handling complex, multi-channel dynamic scenarios, sparking a true revolution in next-generation machine vision, autonomous driving, and in vivo biomedical monitoring.

%

\begin{acknowledgments}	
	
This work was supported by the National Natural Science Foundation of China (Grants No. 12364045, No. 12264028, No. 12304420, and No. U24A20304), the Natural Science Foundation of Jiangxi Province (Grants No. 20232BAB201040, No. 20232BAB211025, and No. 20253BAC260002), the Young Elite Scientists Sponsorship Program by JXAST (Grants No. 2023QT11 and No. 2025QT04).

\end{acknowledgments}


%

\end{document}